\documentclass{article}
\usepackage{authblk} 
\usepackage{graphicx} 
\usepackage[a4paper]{geometry}
\usepackage{graphicx}
\usepackage[textwidth=8em,textsize=small]{todonotes}
\usepackage{natbib}
\usepackage{amsmath}
\usepackage{amsfonts}
\usepackage{setspace}
\usepackage{amsthm}
\usepackage{mdsymbol} 
\usepackage{multirow} 
\usepackage{threeparttable} 
\usepackage{longtable} 
\usepackage{threeparttable}
\usepackage{mathrsfs}
\usepackage[pdfencoding=auto,hidelinks]{hyperref}
\usepackage{graphicx} 
\usepackage{cleveref}
\usepackage{caption}
\usepackage{subcaption} 
\usepackage{natbib} 
\graphicspath{ {./figures/} }

\title{A Bayesian Circadian Hidden Markov Model to Infer Rest-Activity Rhythms Using 24-hour Actigraphy Data}

\author[1]{Jiachen Lu}
\author[2]{Qian Xiao}
\author[3]{Cici Bauer\thanks{corresponding author}}
\affil[1,3]{Department of Biostatistics and Data Science, The University of Texas Health Science Center at Houston School of Public Health}
\affil[2]{Department of Epidemiology, Human Genetics, and Environmental Sciences, The University of Texas Health Science Center at Houston School of Public Health}
\affil[1,2,3]{Center for Spatial‑temporal Modeling for Applications in Population Sciences, The University of Texas Health Science Center at Houston School of Public Health}

\begin{document}

\maketitle

\begin{abstract}

24-hour actigraphy data collected by wearable devices offer valuable insights into physical activity types, intensity levels, and rest-activity rhythms (RAR). RARs, or patterns of rest and activity exhibited over a 24-hour period, are regulated by the body's circadian system, synchronizing physiological processes with external cues like the light-dark cycle. Disruptions to these rhythms, such as irregular sleep patterns, daytime drowsiness or shift work, have been linked to adverse health outcomes including metabolic disorders, cardiovascular disease, depression, and even cancer, making RARs a critical area of health research.

In this study, we propose a Bayesian Circadian Hidden Markov Model (BCHMM) that explicitly incorporates 24-hour circadian oscillators mirroring human biological rhythms. The model assumes that observed activity counts are conditional on hidden activity states through Gaussian emission densities, with transition probabilities modeled by state-specific sinusoidal functions. Our comprehensive simulation study reveals that BCHMM outperforms frequentist approaches in identifying the underlying hidden states, particularly when the activity states are difficult to separate. BCHMM also excels with smaller Kullback-Leibler divergence on estimated densities. With the Bayesian framework, we address the label-switching problem inherent to hidden Markov models via a positive constraint on mean parameters. From the proposed BCHMM, we can infer the 24-hour rest-activity profile via time-varying state probabilities, to characterize the person-level RAR. We demonstrate the utility of the proposed BCHMM using 2011-2014 National Health and Nutrition Examination Survey (NHANES) data, where worsened RAR, indicated by lower probabilities in low-activity state during the day and higher probabilities in high-activity state at night, is associated with an increased risk of diabetes.

\end{abstract}


\clearpage

\section{Introduction}
Biological rhythms are rhythmic patterns inherent in all living organisms at molecular, cellular, physiological, and behavior levels \citep{lamont_circadian_2010}. Of these, circadian rhythms, with an oscillation of approximately 24 hours, are critical within the human body. Other biological rhythms include ultradian and infradian rhythms, which respectively have periods shorter or longer than 24 hours \citep{hobson_cognitive_2002, blum_highly_2014}. A diminished or weakened circadian rhythmicity has been linked to significantly increased risks of metabolic, neurodegenerative, and cardiovascular diseases \citep{montaruli_biological_2021, cheng_predicting_2021, xiao_association_2021, morris_impact_2012}. 

The assessment of circadian rhythmicity has been advanced through established circadian biomarkers such as circulating melatonin level \citep{pandi-perumal_dim_2007, figueiro_effects_2014}. However, such biomarkers require a laboratory standard environment to collect and store the bio-samples from which melatonin levels are measured, yielding only sparse measurements per 24-hour cycle. Alternatively, wearable devices offer the monitoring of circadian fluctuations in a free-living environment. They provide long-time use over multiple days and generate high-resolution data. These wearable devices offer measurements such as body temperature, heart rate, blood pressure and activity movements. The latter, also known as the accelerometer or actigraphy data, have been extensively explored in sleep and circadian research, and demonstrated alignment with standard approaches such as polysomnography (PSG) \citep{ancoli-israel_role_2003, li_novel_2020}.


Hidden Markov models (HMMs) have been extensively applied to feature extraction and classification applications such as speech recognition \citep{gales_application_2007}, bioinformatics \citep{chen_analyzing_2016}, healthcare surveillance systems \citep{luo_bayesian_2021}, and network analysis \citep{mor_systematic_2021,bouguila_hidden_2022}. In actigraphy research, HMMs have been used to characterize the rest-activity cycle \citep{huang_hidden_2018, li_novel_2020, wiggin_covert_2020}, activity modes classification \citep{witowski_using_2014, pober_development_2006}, and identify activity intensity levels \citep{bernard_mixture_2019, chaumaray_mixture_2020, xu_hierarchical_2020}. These applications view actigraphy data as the observations generated by the underlying hidden states via emission distributions, including both parametric (e.g., Gaussian, Poisson and Gamma) and non-parametric approaches (e.g., spline-based models). \cite{witowski_using_2014} demonstrated the HMM-based methods more accurately classify activity modes and energy expenditure levels than traditional cutoff points methods. 
\cite{xu_hierarchical_2020} proposed a hierarchical continuous-time HMM that assumed Poisson emission distributions. In this model, the authors considered six distinct hidden states and zero-inflation was imposed on the state with the lowest mean activity, and subject-level covariates were utilized to model the mean parameters. When applied to the 2003-2006 National Health and Nutrition Examination Survey (NHANES) actigraphy dataset, this model suggested a significant heterogeneity in rest-activity patterns across the population, denoted by the substantial variance in mean activity counts of the same state across different subjects. \cite{li_novel_2020} used an unsupervised machine learning algorithm to identify sleep and wake states from actigraphy data. When juxtaposed with the standard polysomnography (PSG) measurements, their algorithm demonstrated considerable robustness in accurately classifying sleep and wake states. However, \cite{li_novel_2020} also pointed out potential limitations within the two-state model, primarily due to the left-skewed distribution of the observations in the wake state and a heavy-tailed distribution in the sleep state, both influenced by subtle movement variations in behavior during sleep and sedentary periods.

The studies mentioned above largely concentrated on classifying activity modes or intensity levels, while assuming time-invariant transition probabilities. This assumption may fail to effectively capture the temporal dynamics inherent in rest-activity patterns. A recent study by \cite{xiao_rest-activity_2022-1}, with the 2011-2014 NHANES dataset, concluded that rest-activity profiles can be distinctly characterized by both the overall activity amplitude, as well as distinct temporal features such as early-rising and prolonged daytime activity periods. Therefore, it is important and of great interest to characterize the rhythmicity in the rest-activity patterns beyond merely doing classification. Towards this end, \cite{huang_hidden_2018} proposed a novel harmonic HMM incorporating time-varying transition probabilities, building upon the biological mechanism of the circadian rhythm. The empirical oscillatory patterns in the actigraphy data were further explored by \cite{hadj-amar_bayesian_2022} using a Bayesian Hidden Semi-Markov model. While they successfully identified oscillations in the mean activity levels by each state, a comprehensive understanding of individual rest-activity profiles in relation to health outcomes was not investigated.

In this study, we propose a Bayesian Circadian Hidden Markov Model (BCHMM) designed to incorporate 24-hour circadian oscillator reflecting inherent human biological rhythms. The Bayesian inference framework utilizes prior specification to facilitate the identification of the hidden activity states, whilst avoiding substantial computational burden. The proposed BCHMM is implemented using the \texttt{Stan} programming language via the R interface \citep{stan_development_team_rstan_2023}. Our study differs significantly from the majority of previous research (e.g., \cite{witowski_using_2014, hadj-amar_bayesian_2022, xu_hierarchical_2020}) in that our key focus is to use 24-hour actigraphy in characterizing the rest-activity rhythms (RAR), in order to understand how weakened RAR may impact the health outcomes. For this objective, we first apply the BCHMM to the 24-hr actigraphy data to determine the rest-activity profiles, and then investigate these profiles between individuals with diabetes and those without.

The rest of the article is organized as follows. In Section~\ref{sec:motivation}, we introduce the dataset that motivated the proposed work. Section~\ref{sec:method_chmm} discusses the standard HMMs and the incorporation of circadian oscillators. In Section~\ref{sec:method_Bayesian}, we present the Bayesian framework and posterior inference method facilitated by \textit{Stan} programming language. Section~\ref{sec:simulation} includes a comprehensive simulation study to demonstrate the performance of the proposed BCHMM and compare it with the frequentist alternative. In Section~\ref{sec:application}, we illustrate the application of BCHMM on the motivating dataset. R scripts and \textit{Stan} file will be made available on Github.

\section{Motivating Example}\label{sec:motivation}
Our motivating dataset was the National Health and Nutrition Examination Survey (NHANES) study between 2011 and 2014, where actigraphy data were collected 10,000 participants. Participants wore ActiGraph GT3X (by ActiGraph, LLC of Pensacola, FL) continuously for nine days. On the first and last days of this period, data collection typically did not span a full 24 hours, leading to partial observations on these two days. We hence excluded these two days and used the remaining seven days in this analysis.

\begin{figure}[htbp]
    \begin{subfigure}{0.5\textwidth}
    \includegraphics[width=0.9\linewidth, height=4.4cm]{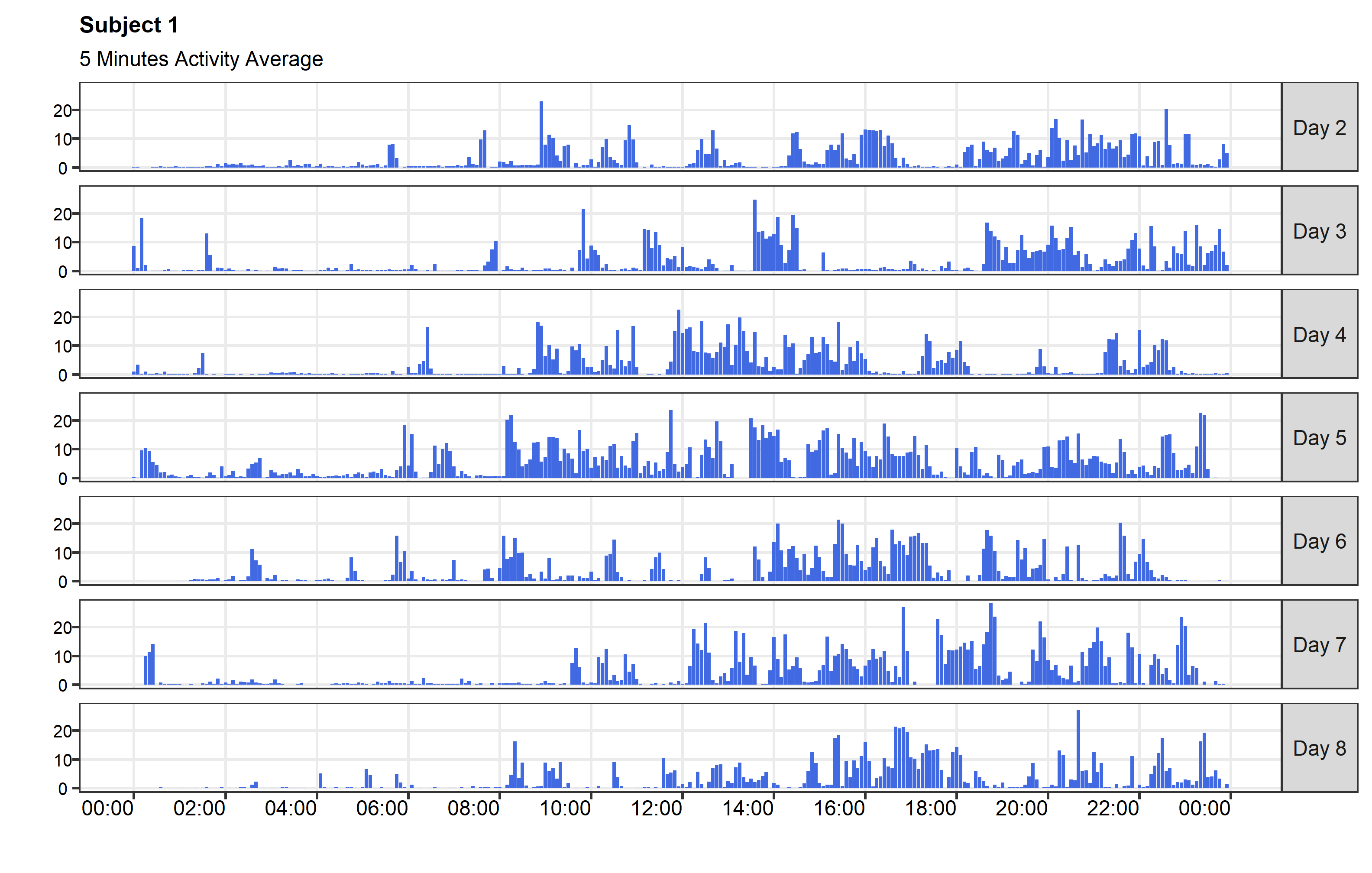} 
    \end{subfigure}
    \hfill
    \begin{subfigure}{0.5\textwidth}
    \includegraphics[width=0.9\linewidth, height=4.4cm]{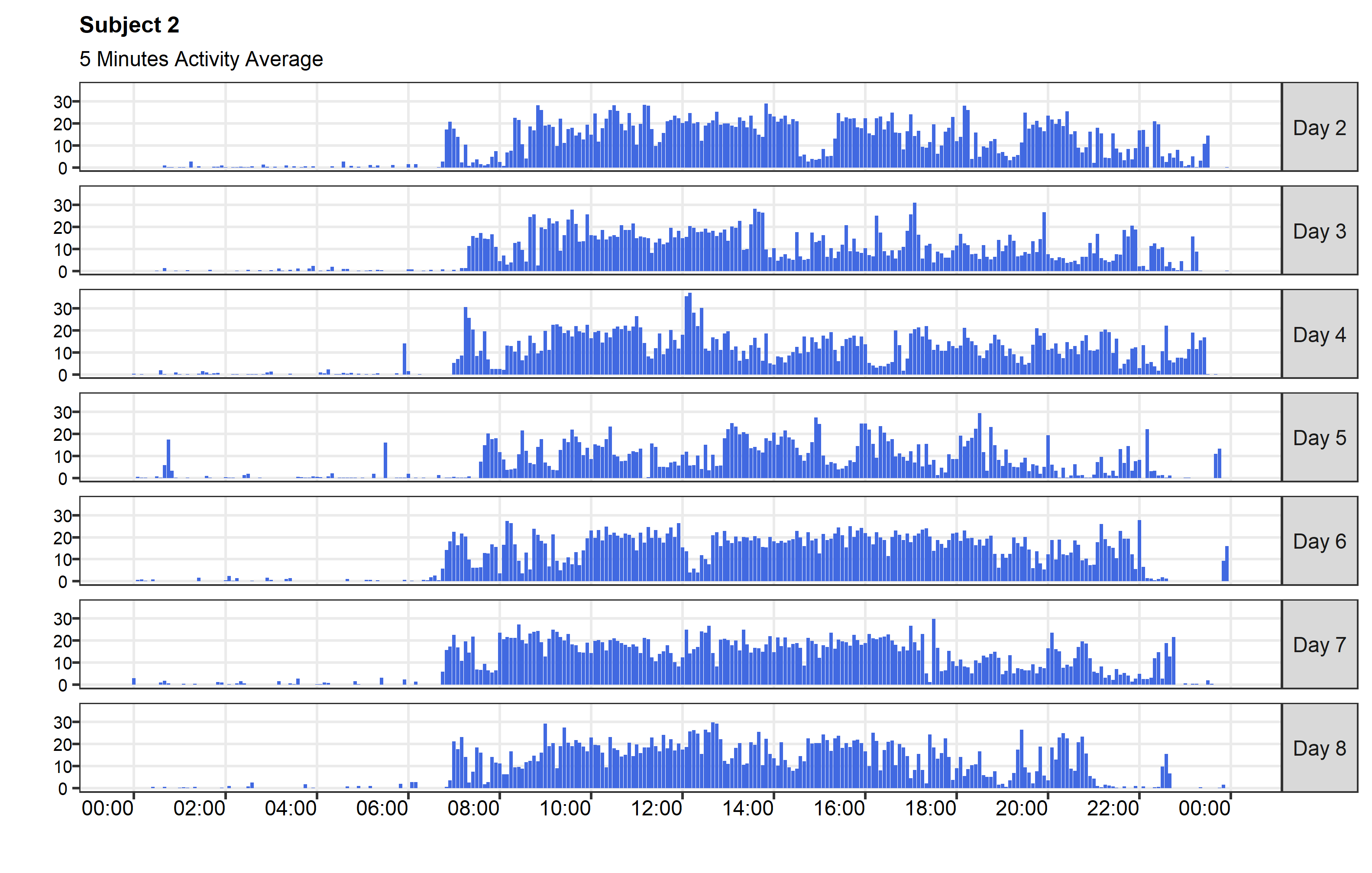}
    \end{subfigure}    

    \caption{Example of the observed 24-hour actigraphy data from two selected subjects in the 2011-2014 NHANES study.}
    \label{fig:raw_act_data}
\end{figure}

The actigraphy device was engineered to capture triaxial accelerations (i.e., $x$-, $y$-, and $z$-axes) at a sampling rate of 80 Hz. The collected data was subsequently summarized on a minute-by-minute basis. Summary measures from triaxial accelerations were then calculated in Monitor Independent Movement Summary (MIMS) units, which is a non-proprietary, open-source, and device-independent universal summary metric~\citep{john_open-source_2019}. In our analysis, we further aggregated the data into 5-min epochs. Examples of the resulting actigraphy data are presented in Figure~\ref{fig:raw_act_data}.

We classified participants' diabetic status based on their Hemoglobin A1c (HbA1c) levels, following the established clinical guidelines for diabetes diagnosis. Specifically, we classified participants as having diabetes (HbA1c $\geq$ 6.5\%), prediabetes (HbA1c between 5.7\% and 6.4\%), normal (HbA1c between 5\% and 5.6\%), or low (HbA1c $<$ 5\%) \citep{palta_hemoglobin_2017}. Here, we focus on diabetes and normal HbA1c groups only, and employed a matched-control study design to select the analytical samples for our analysis objective. Among the adult participants aged over 20, we randomly selected 100 individuals from the diabetes group first. These participants were subsequently matched to a group of individuals with normal HbA1c levels in a 2:1 ratio, stratified by age and gender. The demographic characteristics of the selected samples used for this analysis are provided in Table~\ref{table:table1}. Our goal is to use the model-based approach to extracting meaningful parameters to understand the associations between diabetes risk and weakened rest-activity rhythms (RARs). 

\begin{table}[ht]
\centering
\caption{Summary of the demographics of the analytical sample used in this study.}
\label{table:table1}

\begin{tabular}[t]{llll}
\hline
 & Diabetics & Normal & P-value$^\dag$\\
 & (N=100) & (N=200) & \\
\hline

\multicolumn{4}{l}{\textbf{Age}}\\
\hspace{1em}Mean (SD) & 60.7 (12.2) & 60.3 (11.9) & 0.804\\
\hspace{1em}Median (Min, Max) & 62.0 (21.0, 80.0) & 62.0 (21.0, 80.0) & \\

\multicolumn{4}{l}{\textbf{Age Category}}\\
\hspace{1em}21-30 & 1 (1.0\%) & 2 (1.0\%) & $>$ 0.999\\
\hspace{1em}31-40 & 3 (3.0\%) & 6 (3.0\%) & \\
\hspace{1em}41-50 & 17 (17.0\%) & 34 (17.0\%) & \\
\hspace{1em}51-60 & 24 (24.0\%) & 48 (24.0\%) & \\
\hspace{1em}61-70 & 35 (35.0\%) & 70 (35.0\%) & \\
\hspace{1em}71-80 & 20 (20.0\%) & 40 (20.0\%) & \\

\multicolumn{4}{l}{\textbf{Gender}}\\
\hspace{1em}Female & 48 (48.0\%) & 96 (48.0\%) & $>$ 0.999\\
\hspace{1em}Male & 52 (52.0\%) & 104 (52.0\%) & \\

\multicolumn{4}{l}{\textbf{Race}}\\
\hspace{1em}Mexican American & 17 (17.0\%) & 34 (17.0\%) & $>$ 0.999\\
\hspace{1em}NH Asian & 11 (11.0\%) & 22 (11.0\%) & \\
\hspace{1em}NH Black & 33 (33.0\%) & 66 (33.0\%) & \\
\hspace{1em}NH White & 39 (39.0\%) & 78 (39.0\%) & \\

\hline
\multicolumn{4}{p{11cm}}{$^\dag$Two sample $t$-tests or $\chi^2$ tests are used to test the independence of continuous or discrete variables between diabetes and normal HbA1c groups.} \\
\end{tabular}

\end{table}

\section{Method}\label{sec:method}


In this section, we first briefly review the Hidden Markov Models (HMMs), notations and inference. We then introduce the proposed Bayesian Circadian Hidden Markov Model (BCHMM), and describe its formulation and implementation. 

\subsection{Circadian Hidden Markov Models}\label{sec:method_chmm}
Hidden Markov Model (HMM) is a type of dependent mixture model with two components: an \textit{unobserved} state process denoted by $\{S_t: t \in \mathbb{N}\}$, and \textit{observed} state-dependent process denoted by $\{Y_t: t \in \mathbb{N}\}$. In our analysis, we wish to use the observed individual-level 24-hour actigraphy data to infer the unobserved rest-activity states (such as low-, moderate-, or high-activity states). The observed variable $\boldsymbol y = (y_1, y_2, \dots, y_T)$ is independent of other observations conditional on the current state and has the emission distribution $f$, $y_t|S_t \sim f(\theta_{S_{t}})$. We consider the Gaussian distributions for $f$ where
\begin{equation}\label{eq:gaussion_emission}
    y_t|(S_t = i) \sim N(\mu_{i}, \sigma^2_{i}), i = 1, 2, ..., m, 
\end{equation}
where $\mu_i$s and $\sigma^2_{i}$s are stated-specific mean and variance parameters. Other distributions (e.g., Poisson and Gamma) could also be considered \citep{chaumaray_mixture_2020, witowski_using_2014} as the emission distribution $f$. The unobserved state process is assumed to satisfy the Markov property, where the current state $S_t$ depends on the previous state $S_{t-1}$ via the transition probability matrix $\Gamma=\{\gamma_{ij}\}$, with $\gamma_{ij}=P(S_{t}=j|S_{t-1}=i)$ from state $i$ to $j$, where $i,j = 1,...,m$. The state-specific vector $\boldsymbol{\gamma}_i = (\gamma_{ij})$ is a simplex with $\sum^{m}_{j=1}\gamma_{ij} = 1$. 


To emulate the circadian oscillation inherent in 24-hour actigraphy data, we incorporate a set of harmonic functions as covariates in the transition probabilities, an approach similar to \cite{huang_hidden_2018}. The time-varying transition probability from state $i$ to state $j$ at time point $t$ is then 
\begin{equation}\label{eq:tpm_circadian process}
\begin{split}
    \gamma_{t,ij} &= P(S_{t}=j|S_{t-1}=i) 
    = \frac{\exp(\eta_{t,ij})}{\sum^{m}_{s=1}{\exp(\eta_{t,is})}}, \\
    \eta_{t,ij} &= \beta_{0,ij} + \sum_{l=1}^{L}\Big[
        \beta_{1,ij}^{(l)}\cos\left(2\pi\omega_{l}t \right) + \beta_{2,ij}^{(l)}\sin\left(2\pi\omega_{l}t  \right)\Big],
\end{split}
\end{equation}

\noindent where $l = 1, ..., L$ and $\omega_{l}$ is the inverse of the period parameter. For example, $\omega_{l}=1/24$ is equivalent to a period of 24-hour. These harmonic functions can model additional biological rhythms such as ultradian and infradian rhythms. The circadian coefficients of $\boldsymbol{\beta_{ij}}=(\beta_{0,ij},\beta_{1,ij}^{(l)},\beta_{2,ij}^{(l)})$ are explicitly associated with the transition from state $i$ to state $j$. Larger values of $\eta_{t,ij}$ yield greater transition probabilities. To ensure the identifiability of the $\eta_{t,ij}$, one would impose $\boldsymbol{\beta_{ii}}=\boldsymbol0$. The identifiability of $\boldsymbol{\beta_{ij}}$ is not a concern in our case as long as the entity of interest $\eta_{t,ij}$ is identifiable \citep{wang_bayesian_2023, holsclaw_bayesian_2017}. This model, which we refer to as CHMM, can be implemented using R package \texttt{depmixS4}~\citep{visser_depmixs4_2010} for frequentist inference. 

Non-identifiability of the hidden states concerning the membership of the underlying mixtures is due to the label-switching issue. Specifically, hidden states are arbitrarily labeled such that different orderings of the states can lead to the exact same model \citep{celeux_computational_2000}. Consequently, the model inference can suffer from severe multimodality and substantially increased computational burden. This problem can be addressed by the specification of the priors in a Bayesian framework, as we will discuss in detail next.



\subsection{Bayesian Inference}\label{sec:method_Bayesian}
Our proposed Bayesian Circadian Hidden Markov Models, or BCHMM, can be specified in a two-stage hierarchical form consisting data model and parameter model. The Bayesian framework has several advantages. First, the identifiability of the hidden activity states is facilitated by the specification of appropriate priors. Second, missing data on the observations (e.g., the actigraphy data) could be handled directly via predictive inference in the Bayesian framework. 

The data model is the same as described in Section~\ref{sec:method_chmm}, with the Gaussian emission distributions in equation~\eqref{eq:gaussion_emission} and time-varying transition probabilities in equation~\eqref{eq:tpm_circadian process}. In the parameter model, we specify the priors for the parameters in the data model. We assume the mean parameters of $\boldsymbol{\mu} = (\mu_1, \mu_2, ..., \mu_m)$ independently come from the higher level distribution $\mu_i \sim Gamma(\mu_0, \nu_0)$ with hyperparameters $\mu_0$ and $\nu_0$. The variances $\boldsymbol{\sigma^2} = (\sigma^2_1, \sigma^2_2, ..., \sigma^2_m)$ independently come from scaled inverse-$\chi^2$ distribution, i.e., $\sigma^2_i \sim Inv\chi^2(\kappa_0, \sigma^2_0)$ with hyperparameters $\kappa_0$ and $\sigma^2_0$. The circadian coefficients are specified as $\boldsymbol{\beta_{ij}} \sim N(\mu_\beta, \sigma^2_\beta)$, where $i, j = 1, ..., m$ and $i \neq j$. The initial state probabilities, denoted as $\boldsymbol{\delta} = (\delta_1, ..., \delta_m)^{T}$, follow a flat prior of Dirichlet distribution with concentration parameters $\boldsymbol{\alpha} = (\alpha_1, ..., \alpha_m)$ being all equal.


In our Bayesian framework, label-switching can be fixed by either specifying non-exchangeable priors that are strongly informative or placing ordering constraints on exchangeable priors \citep{betancourt_identifying_2017}. The latter specification results in a non-exchangeable prior as,
\begin{equation*}
    p^{'}(\boldsymbol{\mu}) = 
    \begin{cases} 
          p(\boldsymbol{\mu}), & 0 < \mu_1 < \mu_2 < ... < \mu_m \\
          0, & otherwise 
    \end{cases}
\end{equation*}
Recall that, different orderings of the hidden states can yield the exact same model. 
By placing the ordering constraint on the mean parameters $\mu$s, the mixture likelihood is restricted to be explored under a single ordering rather than arbitrary orderings, such that the inference remains unaffected and the label-switching issue is therefore solved \citep{richardson_bayesian_1997, betancourt_identifying_2017}. Furthermore, it can be easily implemented in \texttt{Stan} by applying \texttt{positive\_ordered} on the mean parameter vectors. 

The marginal likelihood of observations given all parameters $\Theta = (\boldsymbol{\mu}, \boldsymbol{\sigma^2}, \boldsymbol{\beta}, \boldsymbol{\delta})$ can be presented in matrix multiplication as (\cite{zucchini_hidden_2016}) as
\begin{equation}
    L(\boldsymbol{y}|\Theta) = \boldsymbol\delta \boldsymbol{P}(y_{1})\Gamma_{1}\boldsymbol{P}(y_{2})\Gamma_{2} \dots \Gamma_{T-1}\boldsymbol{P}(y_{T})\boldsymbol1^{'}
\end{equation}
where $\boldsymbol{P}(y_{t})$, $t=1, ..., T$ is the emission probability given all possible states, and $\Gamma_t$ is the transition probability matrix at time $t$ with $\Gamma_t = \{\gamma_{t, ij}\}$.
The posterior distribution of $\Theta$ can then be expressed as
\begin{equation}\label{eq:post_distn}
\begin{split}
    p(\Theta|\boldsymbol{y}) 
    \propto & L(\boldsymbol{y}|\Theta)
        p(\boldsymbol{\mu}|\mu_0, \nu_0)
        p(\boldsymbol{\sigma^2}|\kappa_0, \sigma^2_0)
        p(\boldsymbol{\beta}|\mu_{\beta}, \sigma^2_{\beta})
        p(\boldsymbol{\delta}|\boldsymbol{\alpha}) \\
    = & L(\boldsymbol{y}|\Theta) \prod_{i=1}^{m}\Bigg\{
        \Big[\frac{\nu_0^{\mu_0}}{\Gamma(\mu_0)}\mu_i^{\mu_0-1}\text{exp}\big(-\nu_0\mu_i\big)\Big]
        \Big[\frac{(\frac{\kappa_0}{2})^{\frac{\kappa_0}{2}}}{\Gamma(\frac{\kappa_0}{2})}(\sigma_0^2)^{\kappa_0}(\sigma^2_i)^{-(\frac{\kappa_0}{2}+1)}\text{exp}\Big(-\frac{1}{2}\kappa_0(\sigma^2_0)^2\frac{1}{\sigma^2_i}\Big)\Big]
        \\
        & \Bigg[\prod_{j=1, j \neq i}^{m}{
            \frac{1}{\sqrt{2\pi\sigma^2_{\beta}}}\text{exp}\Big(-\frac{(\beta_{ij}-\mu_{\beta})^2}{2\sigma^2_{\beta}}\Big)
            \Bigg]
            }
        {\frac{\delta_i^{\alpha_i-1}}{\Gamma(\alpha_i)}\Gamma\Big(\sum_{i=1}^{m}{\alpha_i}\Big)}
        \Bigg\}
\end{split}
\end{equation}

Inference on parameters relies on integrating out the hidden states which are unknown in the data-generating process. Forward algorithm is used for inference, see details in Supplementary Materials~\ref{sec:supp_forward}. Since the posterior distributions outlined in equation~\eqref{eq:post_distn} are analytically intractable, numerical methods such as Markov Chain Monte Carlo (MCMC) algorithm are required to make inference. We implement the proposed model with \texttt{Stan} programming language \citep{stan_development_team_rstan_2023}, which uses Hamiltonian Monte Carlo (HMC) algorithm and no-U-turn sampler (NUTS) for improved computation efficiency \citep{betancourt_conceptual_2017, neal_mcmc_2011}. Specifically, it uses Hamiltonian dynamics simulation with a Metropolis acceptance step so that proposed values can achieve the targeted distribution more rapidly. The NUTS sampler is an adaptive form of HMC sampling that is implemented in the warm-up steps to optimize parameters in the later HMC iterations. We use the posterior median for inference in the BCHMM and model performance evaluation. 


\section{Simulation Study}\label{sec:simulation}

We conduct an extensive simulation study to assess the performance of the proposed Bayesian Circadian Hidden Markov Model (BCHMM), comparing it with its frequentist counterpart, the CHMM. This simulation design draws inspiration from the 2011-2014 NHANES actigraphy data, taking into account the circadian fluctuations seen in the observed activity counts. The true values used in this simulation, such as the mean and variance of the hidden activity states, have been carefully selected to mirror those found in our preliminary data analysis, ensuring that the results offer relevant insights into the application of the proposed model for the analysis of 24-hour actigraphy data.

\subsection{Simulation Setup}
\begin{figure} [htbp]
    \centering
    \includegraphics[width=.9\linewidth]{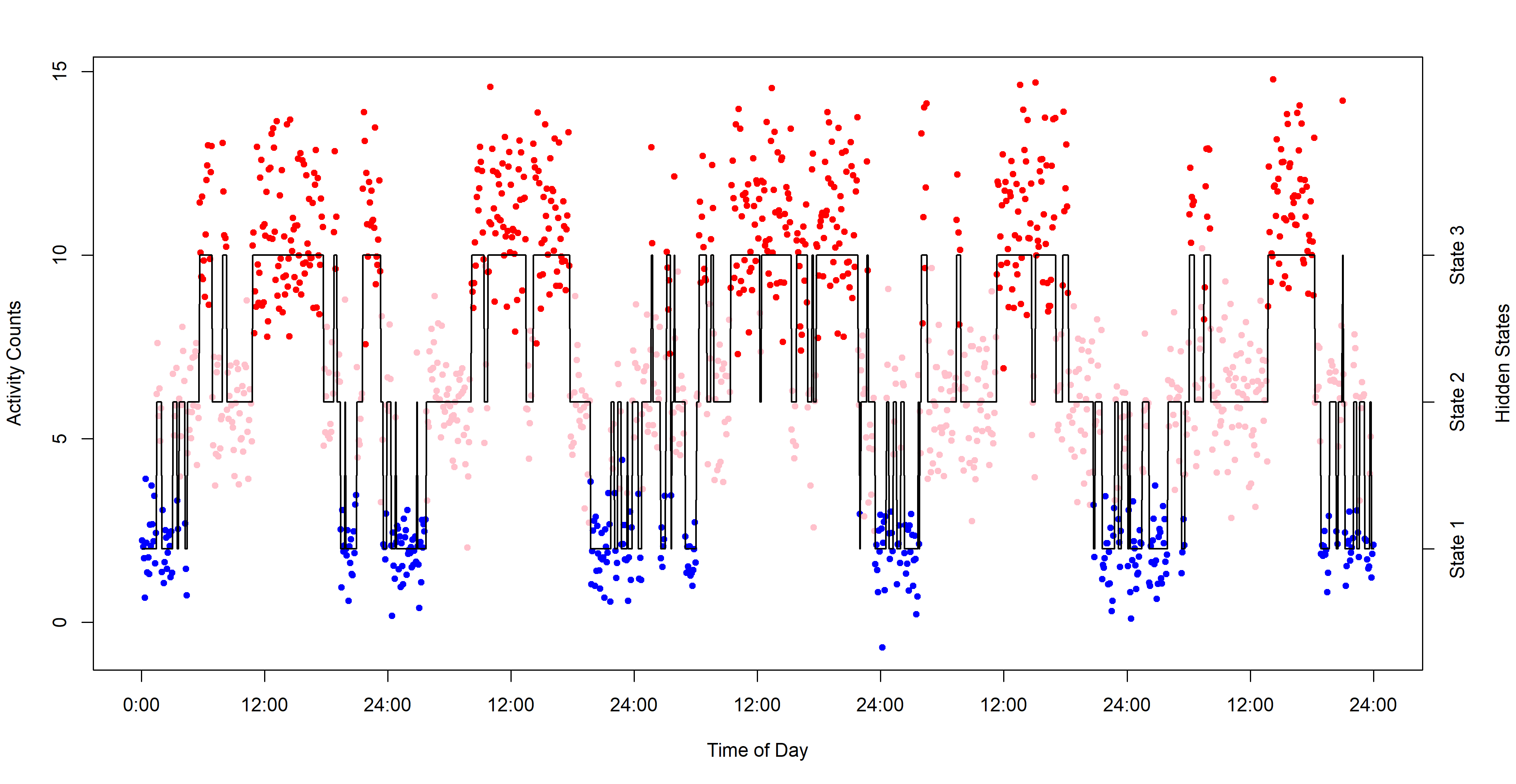} 
    \caption{Illustration of the simulated hidden activity states (black line) and the associated observed activity counts (color-coded dots). Simulated observations from the low-activity (state 1), moderate-activity (state 2), and high-activity (state 3) states are presented in blue, pink, and red colors, respectively.}
  \label{fig:simulated_data}
\end{figure}

We consider three scenarios, each with identical circadian coefficients $\boldsymbol{\beta}$ yet differing underlying hidden states and Gaussian emission distributions. Table~\ref{table:true parameters} presents true parameter values used in this simulation study. In all scenarios, we assume three hidden activity states of low (state 1), moderate (state 2), and high (state 3) activity, and introduce greater overlap between the underlying hidden states from scenarios 1 to 3 by increasing the variance. The mean values are chosen based on the existing literature \citep{huang_hidden_2018, li_demographic_2021} and the empirical distribution of the actigraphy data from the analytical dataset described in Section~\ref{sec:motivation}, with a sample median of 5.5 (interquartile range [IQR] 0.5, 14.6) MIMS/min. Therefore, we choose the mean values of 2, 6, and 11 for low-, moderate- and high-activity states, respectively. In Scenario 1, we assume equal variability of actigraphy values across hidden states (i.e., $\sigma^2_i = 0.5, i=1, 2, 3$). In Scenario 2, we increase the variability by employing a larger value of $\sigma^2_i$ to increase the overlapping of the Gaussian emission distributions. In Scenario 3, the variability within the hidden states increases with the activity level, creating further overlap between the hidden states. Based on the assumed values of circadian coefficients $\boldsymbol{\beta}$ (Supplementary Table~\ref{table:simu_true_coef}), we back-calculate the time-varying transition probability matrices $\Gamma_t=\{\gamma_{t,ij}\}$ according to equation~\eqref{eq:tpm_circadian process}. We simulate the participant-level actigraphy data for 5 consecutive days at 5-min epoch per day, resulting in a total of $T=5\times 288=1440$ observations per participant. For each scenario, we generate 100 participants. Figure~\ref{fig:simulated_data} presents an example of such simulated actigraphy data. 


For comparison, we apply both the proposed BCHMM and CHMM to the simulated data, with CHMM implemented using R package \texttt{depmixS4} \citep{visser_depmixs4_2010}. We run BCHMM with two MCMC chains and monitor every 5,000 iterations for convergence check. We run each chain for 20,000 iterations with the first 10,000 discarded as the burnins, and collect every 4th sample for a total of 2,500 samples. The convergence is assessed by trace plots, a split-$\hat{R}$ statistic of less than 1.05 and a ratio of effective size greater than 0.5 \citep{vehtari_rank-normalization_2021, gelman_bayesian_nodate}. Example trace plots are provided in the Supplementary Figures~\ref{fig:traceplot_means_vars} and \ref{fig:traceplot_betas}.



\begin{table}[htbp]
\centering
\caption{True values of the mean and variance parameters of the hidden activity states used in the simulation study.}
\label{table:true parameters}

\begin{tabular}{l|l|lll}\hline
      Scenario & Parameters   & low-activity & moderate-activity &high-activity \\
               & &  state 1 & state 2 & state 3       \\ \hline
      Scenario 1 &   $\mu$ & 2.0 & 6.0 & 11.0      \\
                 &   $\sigma^{2}$ & 0.5 &0.5 & 0.5   \\  
      Scenario 2 &   $\mu$ & 2.0 & 6.0 & 11.0      \\
                 &   $\sigma^{2}$ & 1.0 &1.0 & 1.0   \\ 
      Scenario 3 &   $\mu$ & 2.0 & 6.0 & 11.0      \\
                 &   $\sigma^{2}$ &0.6 &1.75 & 2.2 \\
     \hline
\end{tabular}
\end{table}

For model checking, we evaluate the parameter estimation of the emission distributions (i.e., $\mu_i$ and $\sigma^2_i$), and the time-varying state probabilities $\boldsymbol{P(S_t)}=(P(S_t=1), P(S_t=2), P(S_t=3))$, a three-dimensional vector of probabilities for a 24-hour period at 5-min epoch (i.e., $t=1, 2, \dots, 288$). To get the posterior inference of $\boldsymbol{P(S_t)}$, we first obtain get the time-varying transition probabilities $\hat{\Gamma}_t$ via equation~\eqref{eq:tpm_circadian process}. The estimated $\boldsymbol{P(S_t)}$ can then be obtained by the product of the estimated initial probabilities $\boldsymbol{\hat{\delta}}$ and time-varying transition probabilities $\boldsymbol{\hat{\Gamma}_t}$ from 

\begin{equation}\label{eq:sp_bias}
    \boldsymbol{\hat{P}(S_t)} = \boldsymbol{\hat{\delta}}\hat{\Gamma}_1 ... \hat{\Gamma}_t.
\end{equation}

To evaluate and compare model performance, we measure the following metrics: mean absolute bias (MAB), mean of mean squared error (MMSE), mean of posterior standard deviation (MSD), and mean coverage rate (MCR). Here, $\theta$ denotes the parameter of interest, $\hat{\theta}$ the estimates either from BCHMM or CHMM, and $\theta_{True}$ the true value: 
\begin{equation}\label{eq:simualtion_metrics}
    \begin{split}
    \text{MAB}_{\theta} &= 
    \frac{1}{R}\sum_{r=1}^{R}|\hat{\theta}-\theta_{True}|, \\
    \text{MMSE}_{\theta} &= 
    \frac{1}{R}\sum_{r=1}^{R}(\hat{\theta}-\theta_{True})^2, \\
    \text{MSD}_{\theta} &= 
    \frac{1}{R}\sum_{r=1}^{R}\Big(\frac{1}{N_{iter}}\sqrt{Var_{\theta \sim p(\theta|\boldsymbol{y})}(\theta)} \Big), \\
    \text{MCR}_{\theta} &= 
    \frac{1}{R}\sum_{r=1}^{R} I\{\hat{\theta} \in 95\%CP\}. \\
    \end{split}
\end{equation}
The metrics are calculated across a set of replicates $r$, with $R$ being the total number of these replicates (i.e., $R=100$). 
For MSD, $N_{iter}$ is the number of iterations in the MCMC sampling and $p(\theta|\boldsymbol{y})$ is the posterior density for $\theta$. For MCR, $I\{\}$ is the indicator function with value 1 when $\hat{\theta}$ is within the $95\%$ coverage probability ($95\%CP$). For time-varying state probabilities, the mean 24-hour accumulated absolute bias is further evaluated, defined as,
\begin{equation}\label{eq:acc_sp_bias}
    \text{MAB}_{P(S_t)} = \frac{1}{R}\sum_{r=1}^{R}\sum_{t=1}^{T_{24h}}|\boldsymbol{\hat{P}(S_t)} - \boldsymbol{P_{True}(S_t)}|
\end{equation}
\noindent where $T_{24h}=288$ stands for a 24h period.



\subsection{Simulation Results}

\begin{figure}[htbp]
    \centering
    \includegraphics[width=.8\linewidth]{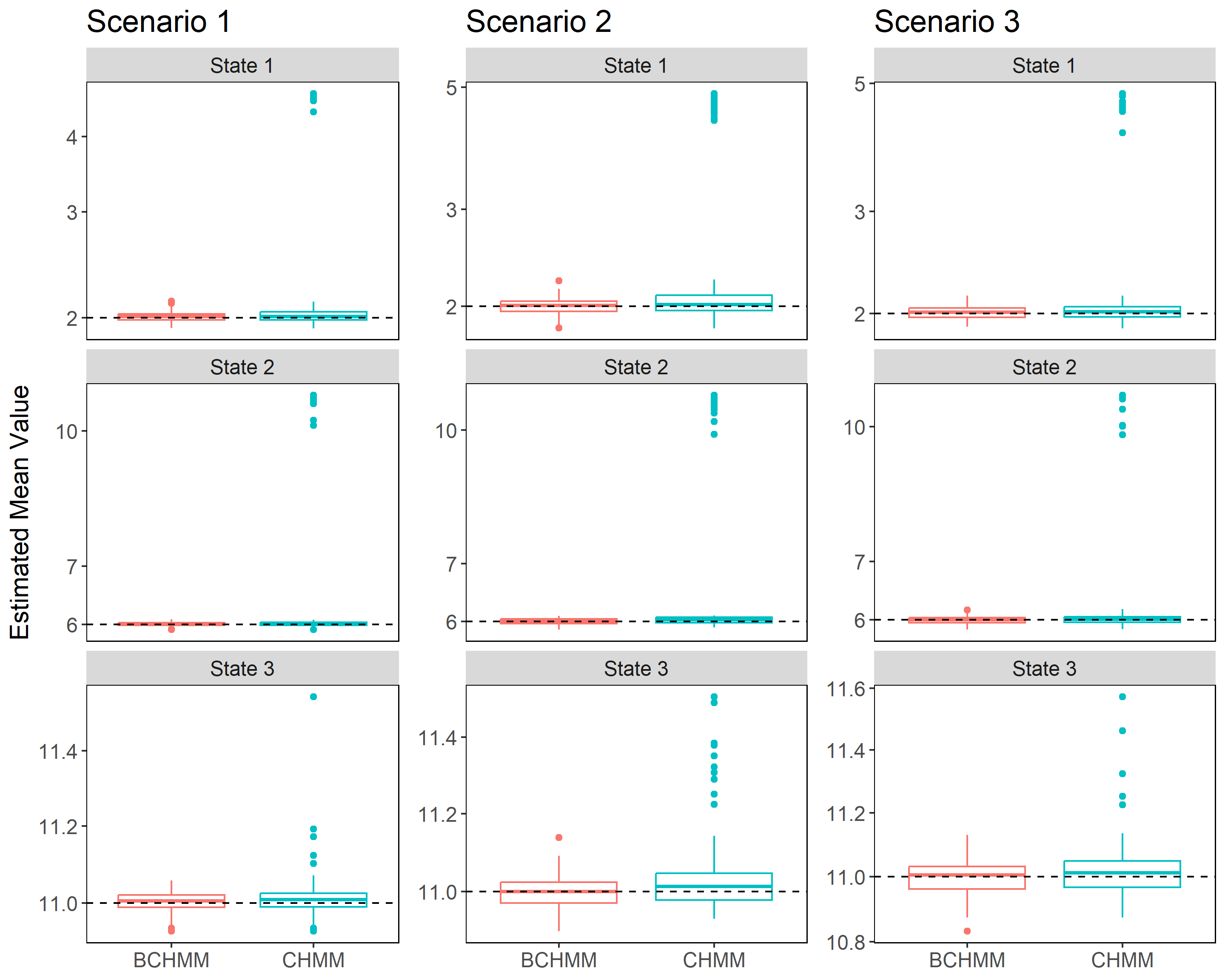} 
    \caption{Comparison of estimated state means from BCHMM and CHMM. True values are indicated in black dashed lines.}
    \label{fig:mean_chmm_vs_bchmm}

\end{figure}

\begin{figure}[htbp]
    \begin{subfigure}{0.5\textwidth}
    \includegraphics[width=1\linewidth]{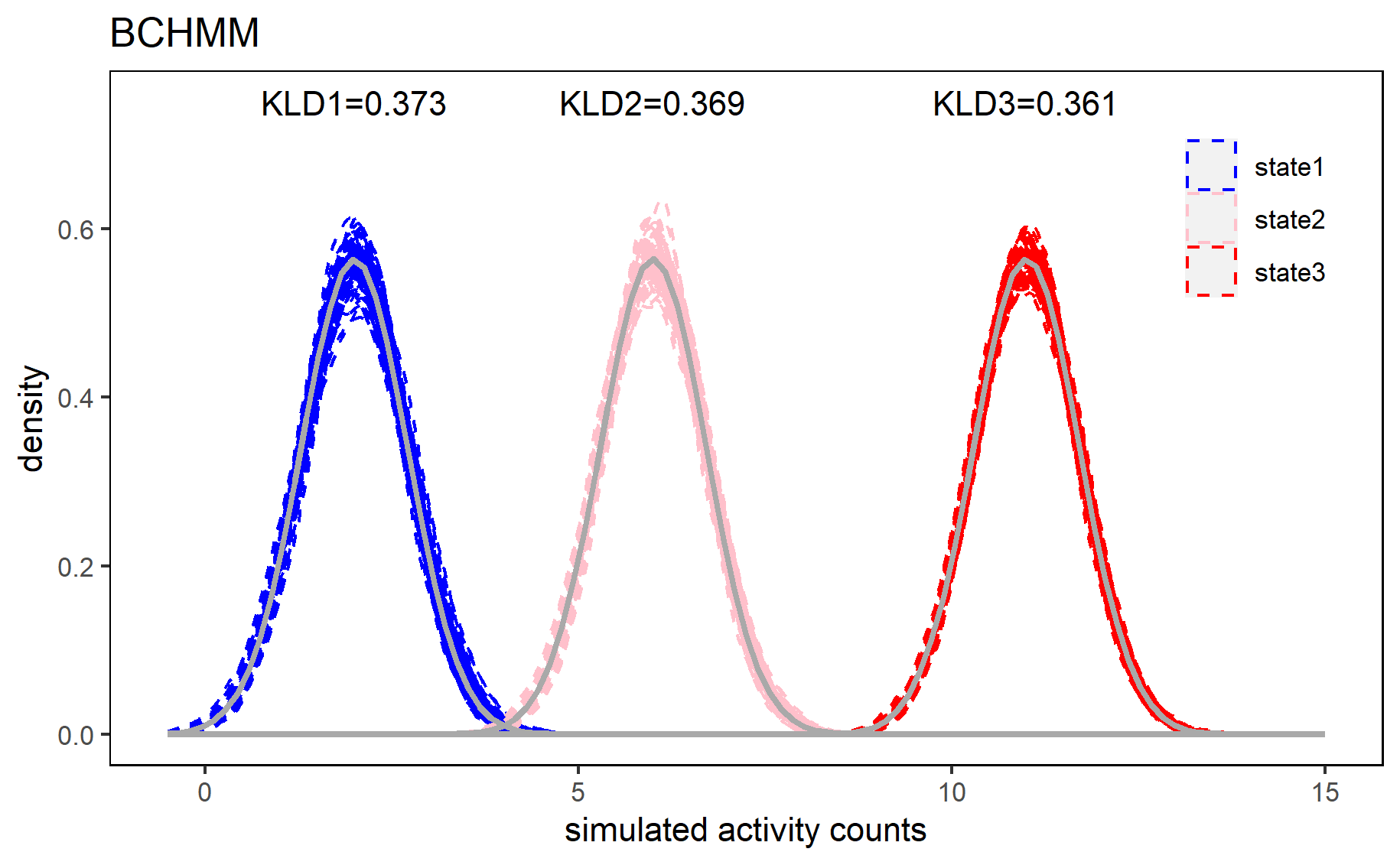} 
    \end{subfigure}
    \hfill
    \begin{subfigure}{0.5\textwidth}
    \includegraphics[width=1\linewidth]{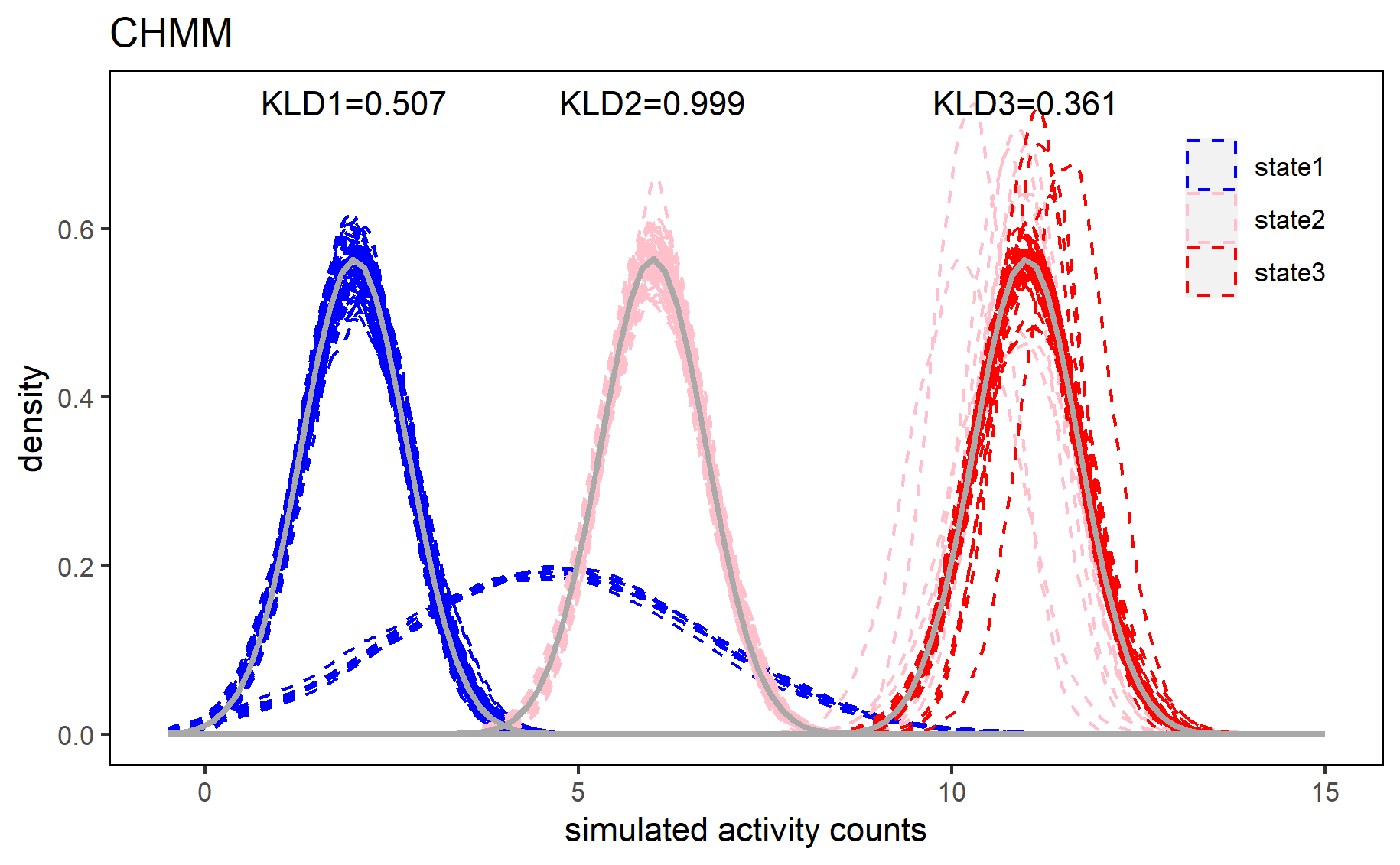}
    \end{subfigure}    

    \begin{subfigure}{0.5\textwidth}
    \includegraphics[width=1\linewidth]{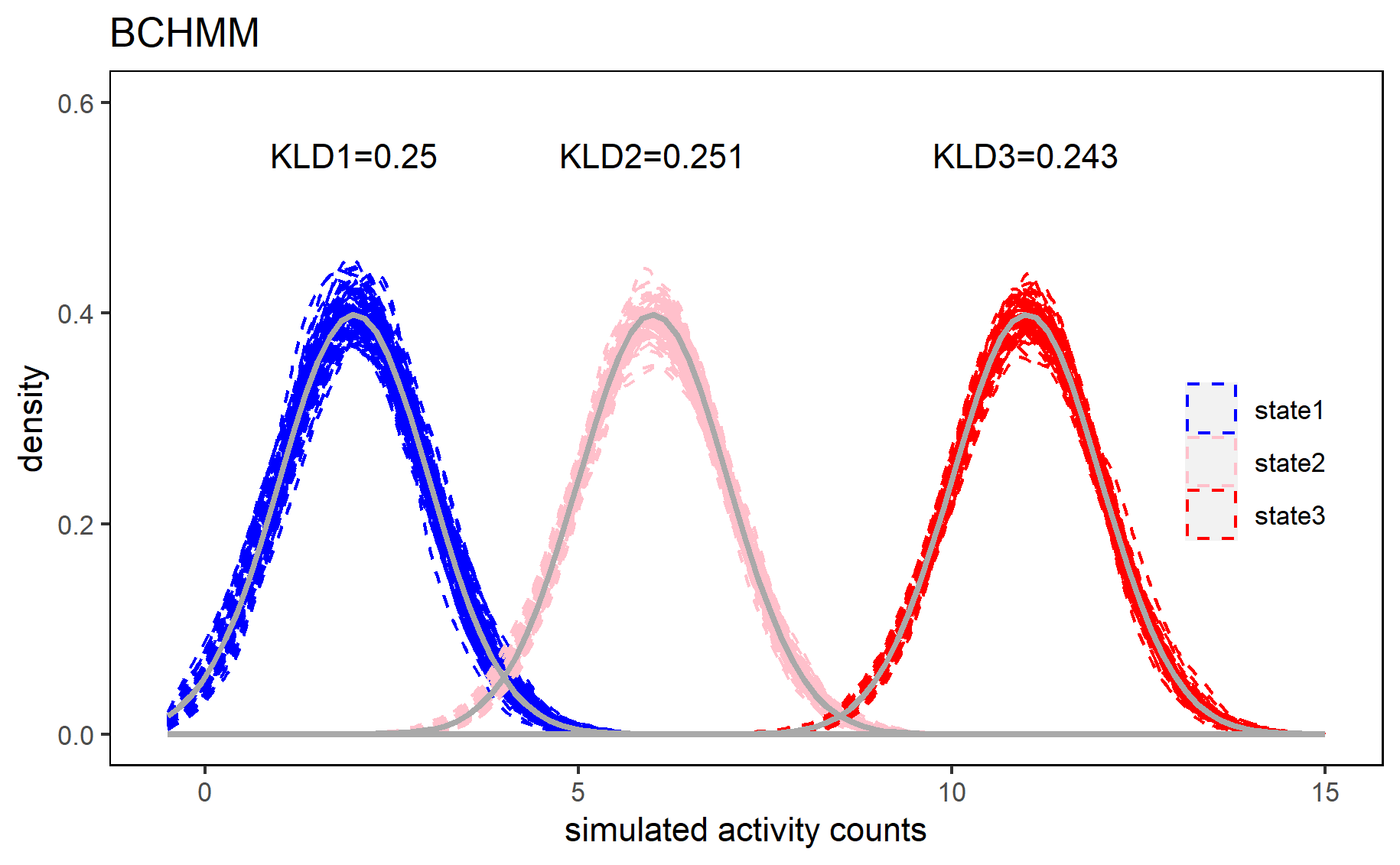} 
    \end{subfigure}
    \hfill
    \begin{subfigure}{0.5\textwidth}
    \includegraphics[width=1\linewidth]{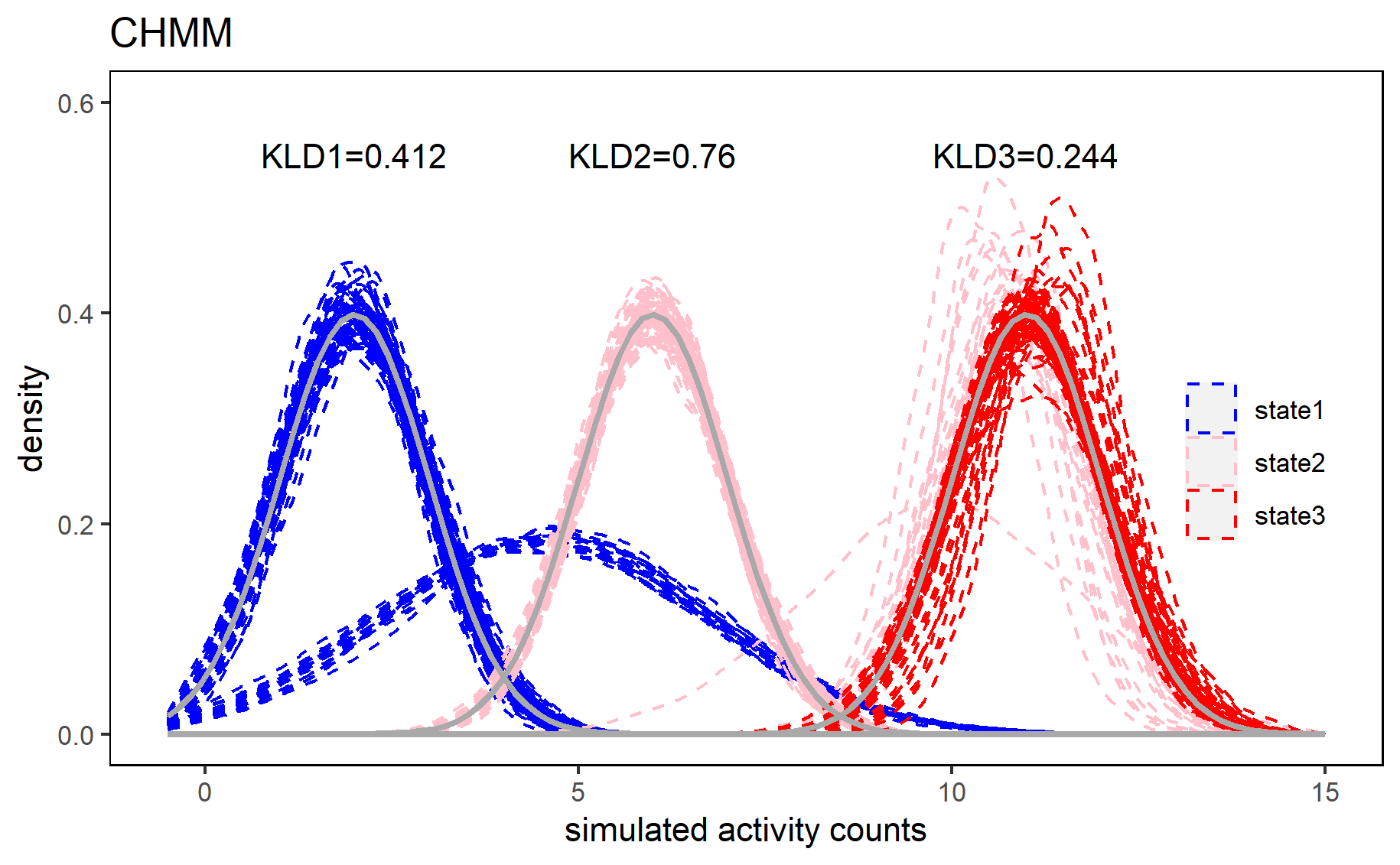}
    \end{subfigure}    

    \begin{subfigure}{0.5\textwidth}
    \includegraphics[width=1\linewidth]{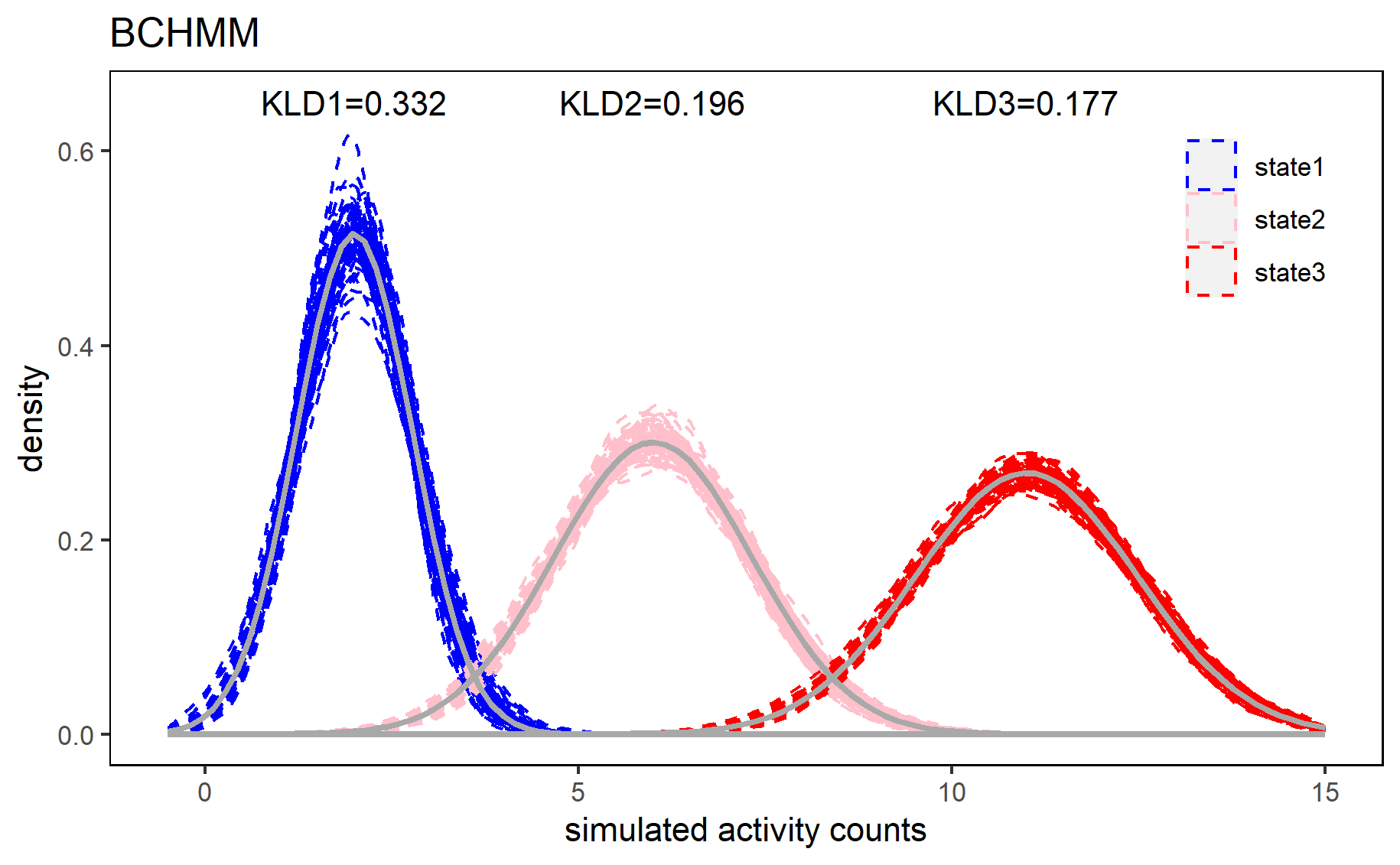} 
    \end{subfigure}
    \hfill
    \begin{subfigure}{0.5\textwidth}
    \includegraphics[width=1\linewidth]{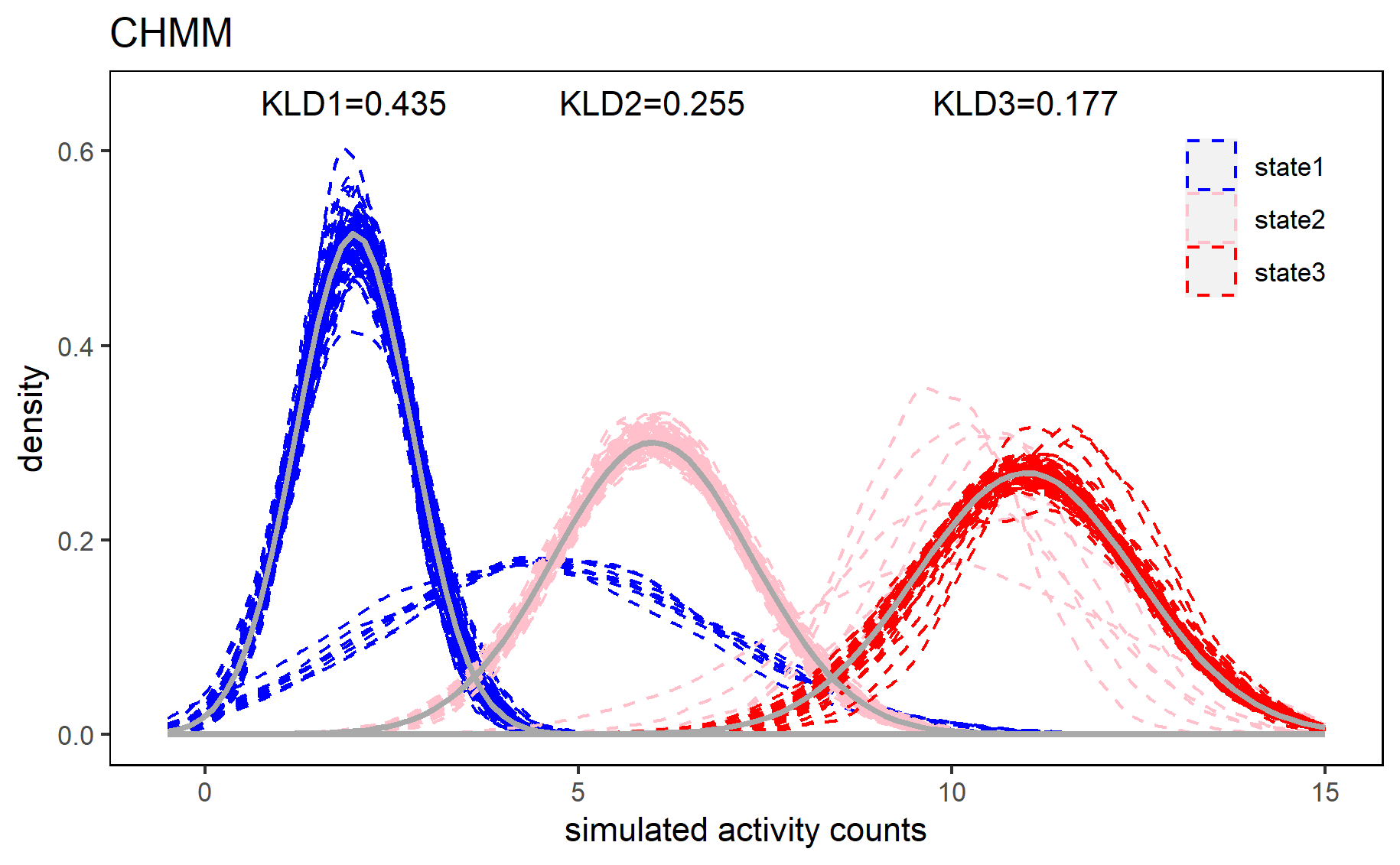}
    \end{subfigure}    

    \caption{Estimated Gaussian emission densities from three hidden activity states. Grey lines are true densities. Estimated densities are color-coded as blue for state 1, pink for state 2 and red for state 3, one for each replicate.}
    \label{fig:simulation_density_all_scenarios}
\end{figure}

In all scenarios, BCHMM generally outperforms CHMM in accurately identifying the underlying hidden states (Figure~\ref{fig:mean_chmm_vs_bchmm} and Figure~\ref{fig:simulation_density_all_scenarios}). Particularly, CHMM tends to overestimate state means $\mu_i$ as mixing the hidden states 1 and 2 and misclassifying state 2 as state 3, a pattern not seen in BCHMM estimates. This overestimation becomes more severe as overlap increases in underlying Gaussian distributions in scenarios 2 and 3, as evidenced by outliers in CHMM estimates in Figure~\ref{fig:mean_chmm_vs_bchmm} boxplots. BCHMM also has smaller MAB compared to CHMM, especially for states 1 and 2 (Table~\ref{table:bchmm_vs_chmm_results}). Parameter estimates from BCHMM are also more reliable and robust, with MABs less than 0.09 and the standard deviations on the absolute bias below 0.08 across all states and scenarios (Table~\ref{table:bchmm_vs_chmm_results}). Higher values of Kullback-Leibler Divergence (KLD) indicate larger departures from true values, as seen in CHMM. For example, the KLD for the estimated state 2 from CHMM is 0.76, considerably higher than the 0.251 for BCHMM (pink lines in the middle panel in Figure~\ref{fig:simulation_density_all_scenarios}).

\begin{table}[htbp]
\centering
\caption{Comparison of accuracy on estimated parameters for BCHMM and CHMM from the simulation study. Results presented in the table are the MAB, and standard deviation of MAB in parentheses.}
\label{table:bchmm_vs_chmm_results}

\begin{tabular}[t]{c p{2cm}p{2cm}p{2cm}p{2cm}p{2cm}p{2cm}}
\hline
\multicolumn{1}{c}{ } & \multicolumn{2}{c}{Scenario 1} & \multicolumn{2}{c}{Scenario 2} & \multicolumn{2}{c}{Scenario 3} \\
Parameter & BCHMM & CHMM & BCHMM & CHMM & BCHMM & CHMM\\
\hline
$\mu_1$ & 0.033 (0.026) & 0.267 (0.749) & 0.056 (0.045) & 0.534 (0.997) & 0.040 (0.027) & 0.243 (0.687)\\

$\mu_2$ & 0.026 (0.020) & 0.451 (1.36) & 0.038 (0.024) & 0.925 (1.84) & 0.048 (0.038) & 0.392 (1.18)\\

$\mu_3$ & 0.021 (0.016) & 0.033 (0.060) & 0.032 (0.025) & 0.067 (0.103) & 0.044 (0.036) & 0.062 (0.084)\\

$\sigma^2_1$ & 0.033 (0.026) & 0.368 (1.06) & 0.075 (0.052) & 0.769 (1.44) & 0.044 (0.040) & 0.398 (1.20)\\

$\sigma^2_2$ & 0.028 (0.022) & 0.038 (0.046) & 0.055 (0.042) & 0.102 (0.233) & 0.108 (0.070) & 0.174 (0.389)\\

$\sigma^2_3$ & 0.018 (0.013) & 0.026 (0.036) & 0.048 (0.037) & 0.075 (0.086) & 0.098 (0.080) & 0.119 (0.129)\\
\hline
\end{tabular}

\end{table}

%
%

Regarding the state probabilities, BCHMM has a smaller median accumulated bias (i.e. $\text{MAB}_{P(S_t)}$) and less extreme bias compared to CHMM (Table~\ref{table:bchmm_vs_chmm_results_sp} and Supplementary Figure~\ref{fig:simulation_sp_acc_bias}). The range of $\text{MAB}_{P(S_t)}$ is also much narrower in BCHMM than in CHMM. True state 1 probabilities decrease from midnight to 8 am and increase from 8 pm to midnight, remaining low throughout the day. Yet, CHMM yields consistently higher estimates for state 1 throughout the day, likely due to its overestimation of the state 1 mean. Conversely, BCHMM recovers the true state better than CHMM, with less bias at each time point across all states and scenarios. The median estimated state probabilities from BCHMM (depicted by red solid lines) align more closely with the true values (black solid line), with narrower coverage bands across all 100 replicates (see Supplementary Figures~\ref{fig:simulation_est_sp_bias} (a), (c), and (e)).

\begin{table}[ht]\centering
\caption{Comparison of 24-hour accumulated absolute bias on estimated state probabilities for BCHMM and CHMM. Results are summarized as the median (min, max) of the accumulated absolute bias defined in Equation~\eqref{eq:acc_sp_bias}.}
\label{table:bchmm_vs_chmm_results_sp}

\begin{tabular}[t]{p{1.1cm} p{1.9cm} p{2.2cm} p{1.9cm} p{2.2cm} p{1.9cm} p{2.2cm}}
\hline
\multicolumn{1}{c}{ } & \multicolumn{2}{c}{Scenario 1} & \multicolumn{2}{c}{Scenario 2} & \multicolumn{2}{c}{Scenario 3} \\
 & BCHMM & CHMM & BCHMM & CHMM & BCHMM & CHMM\\
\hline
State 1 & 10.08 & 11.21 & 8.90 & 10.77 & 9.79 & 10.84\\
& (1.74, 24.14) & (2.63, 121.76) & (1.20, 21.85) & (1.01, 119.77) & (2.60, 23.78) & (3.08, 114.47)\\

State 2 & 17.36 & 18.69 & 16.55 & 18.18 & 16.87 & 17.80\\
 & (4.87, 32.57) & (5.43, 94.06) & (4.43, 32.39) & (4.42, 119.63) & (6.10, 31.38) & (6.51, 85.17)\\

State 3 & 16.04 & 18.01 & 16.39 & 18.51 & 14.78 & 16.94\\
 & (3.92, 38.20) & (5.04, 94.54) & (5.42, 36.98) & (4.60, 119.46) & (3.87, 37.27) & (3.61, 79.91)\\
\hline
\end{tabular}

\end{table}

\section{Motivating Example Revisited}\label{sec:application}
We now revisit the 2011-2014 NHANES motivating example introduced in Section~\ref{sec:motivation} and apply the proposed BCHMM to the NHANES actigraphy dataset. Similar to the approach adopted in \cite{huang_hidden_2018} and \cite{li_novel_2020}, we apply a square-root transformation to the observed actigraphy data and assume three hidden states of low-, moderate- and high-activity states. Weakly informative priors are considered for other parameters, with $\mu_i \sim Gamma(1, 1)$, $\sigma_i^2 \sim Inv\chi^2(2, 0.5)$ and $\beta_{ij} \sim N(0, 10)$.  


In addition to the model parameters introduced in BCHMM earlier, we also compare the derived rest-activity rhythm measures proposed by \cite{huang_hidden_2018}. We give a brief overview here but encourage interested readers to refer to the original article. Our primary interest is the \textit{Rhythmic Index} (RI), a measure indicating overall rhythmicity on a scale from 0 with lower values for worsened rhythmicity. RI is computed as,
\begin{equation}
    RI=\frac{24}{24-a}\Big(
    \frac{1}{a}\int_{t\in I_c}P(S_t=1)dt - \frac{a}{24}
    \Big)
\end{equation}
where $S_t=1$ indicates the low-activity state and $I_c = [c-a/2, c+a/2]$ suggests the low-activity window under perfect rhythmicity. Here the gravity center of the low-activity state window is denoted as $c$. The total amount of rest, denoted as $a$, is estimated by the total length in the low-activity state during a full day of 24-hour, and calculated as $\int_{t\in T_{24hr}}P(S_t=1)dt$. We also fit the CHMM model to the motivating example for comparison. 

\begin{figure}[htbp]
    \centering
    \includegraphics[width=.9\linewidth]{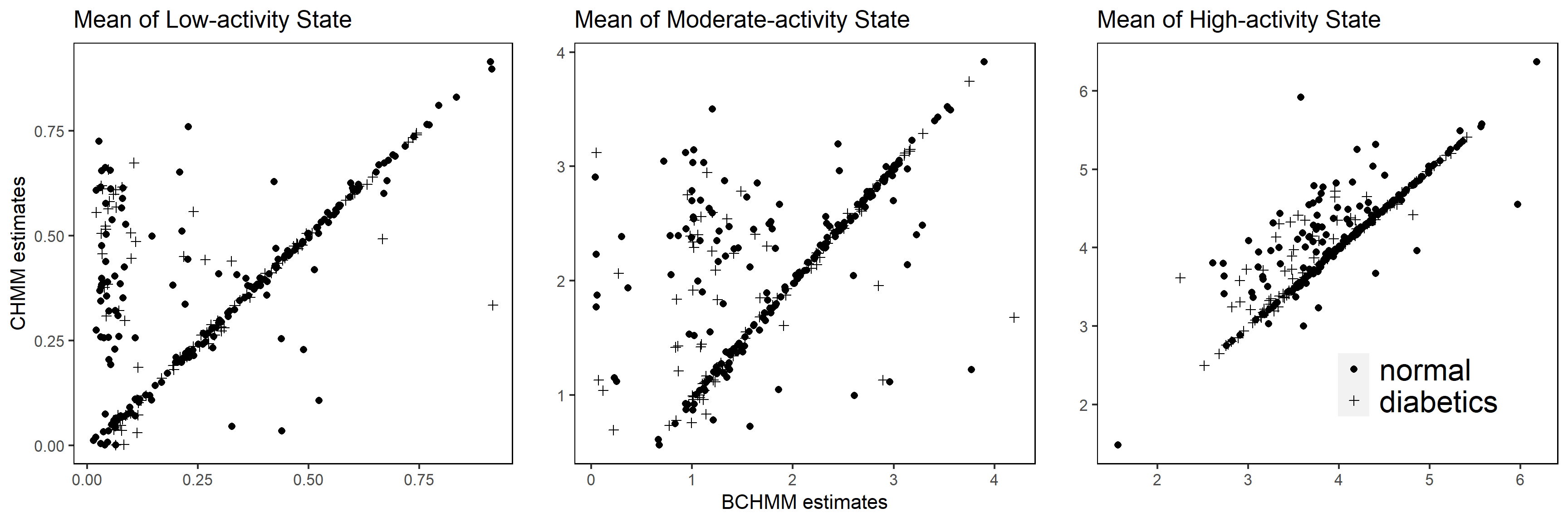} 
    \caption{Comparison of estimated state means from BCHMM and CHMM on NHANES subjects. BCHMM estimates are on $X$ axis while CHMM estimates on $Y$ axis, with the solid dot symbols representing diabetics and the plus symbol representing the normal HbA1c group.}
    \label{fig:mean_chmm_vs_bchmm_nhanes}

\end{figure}

Figure~\ref{fig:mean_chmm_vs_bchmm_nhanes} presents the estimated activity state means $\mu$s from the BCHMM and CHMM. Each point corresponds to an individual participant, with estimates from BCHMM and CHMM presented by the $X$ and $Y$ axes, respectively. Compared to BCHMM, CHMM tends to overestimate the mean values of the hidden states, as evidenced by points concentration on the top left corner above the diagonal line. The extent to which CHMM overestimates the mean is particularly severe for the low-activity state, consistent with the findings from the simulation study that CHMM fails to distinguish the low- and moderate-activity hidden states and hence overestimates the means. The estimated variance from CHMM is also substantially larger than that from BCHMM (Supplementary Figure~\ref{fig:var_chmm_vs_bchmm_nhanes}).

\begin{figure}[htbp]
    \centering
    \includegraphics[width=.9\linewidth]{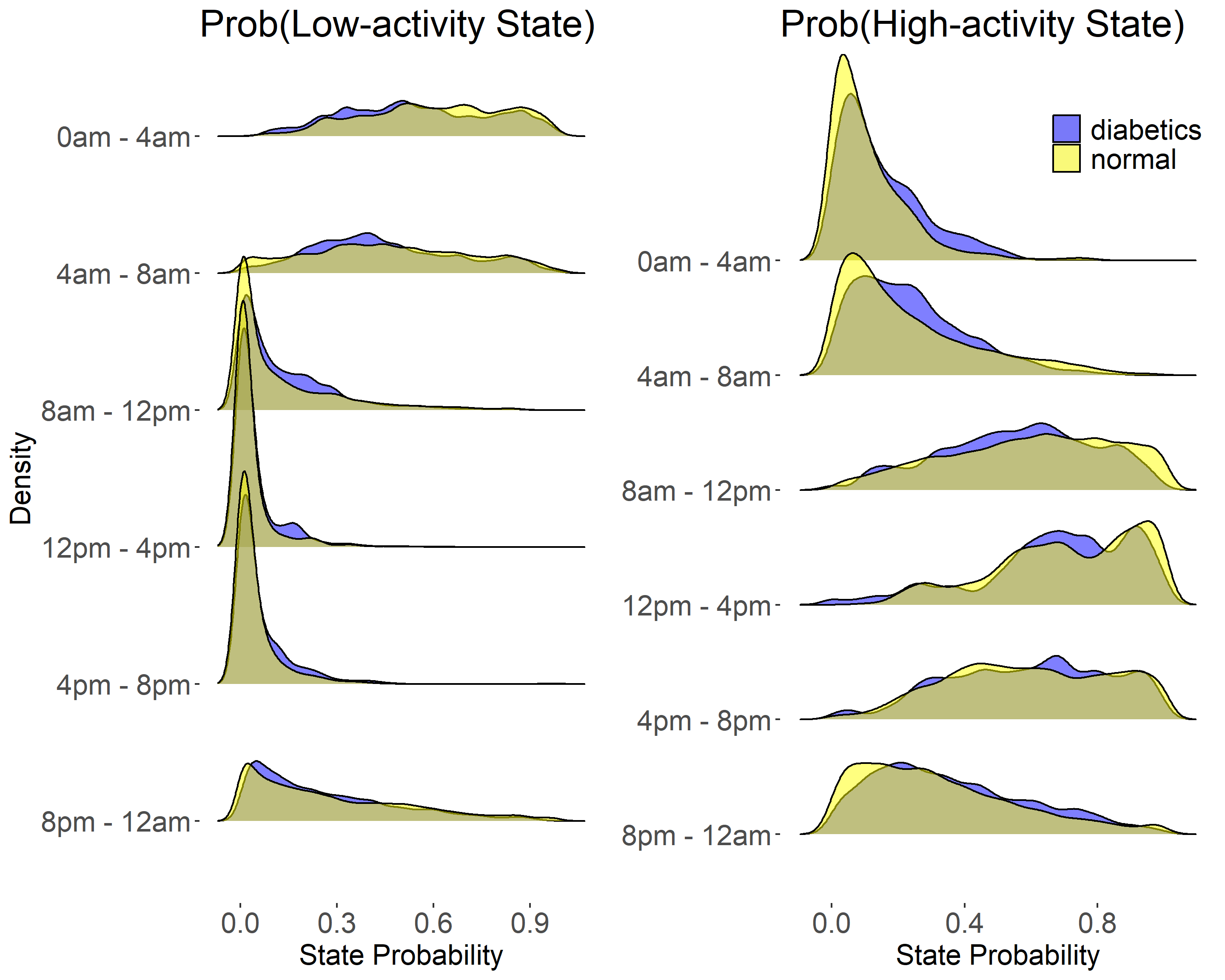} 
    \caption{BCHMM-estimated state probabilities by diabetic status over a 24-hour period, segmented into 4-hour intervals (rows). For each time window, the figure compares the density distributions in the low-activity state (left panel) and high-activity state (right panel), color-coded by subjects with diabetes and normal HbA1C.}
    \label{fig:sp1_sp3_bchmm_nhanes}

\end{figure}

Figure~\ref{fig:sp1_sp3_bchmm_nhanes} presents the estimated time-varying state probabilities from BCHMM, comparing the participants with diabetes and those without. The distribution of probabilities for the low-activity and high-activity states are compared by 4-hour segments over a 24-hour period. Differences between the two groups can be clearly discernible in the non-overlapping, color-shaded areas, where blue and yellow represent the diabetic and normal HbA1c groups, respectively. Especially evident during the daytime (8 am-12 pm and 12-4pm), individuals with diabetes (blue) demonstrate a higher likelihood of being in the low-activity state and a lower likelihood of high-activity state compared to the normal group (yellow). However, this trend is inverted during the nighttime hours (0-4 am). In essence, the temporal patterns inferred from the time-varying state probabilities suggest that individuals with diabetes tend to be more active during the night and less active during the day compared to their normal counterparts. On average, diabetic precipitants have a mean activity count of 3.8 for the high-activity state, equivalent to 72.2 on the original scale of 5-minute aggregation. This is significantly lower than the 4.05 observed for the normal group (82.0 on the 5-minute aggregation scale), suggesting that individuals in the diabetic group are typically less active. Report of additional parameters are summarized in Supplementary Table~\ref{table:nhanes_bchmm_chmm_est_RI}. 

\begin{figure}[htbp]
    \begin{subfigure}{0.5\textwidth}
    \includegraphics[width=0.9\linewidth, height=4.4cm]{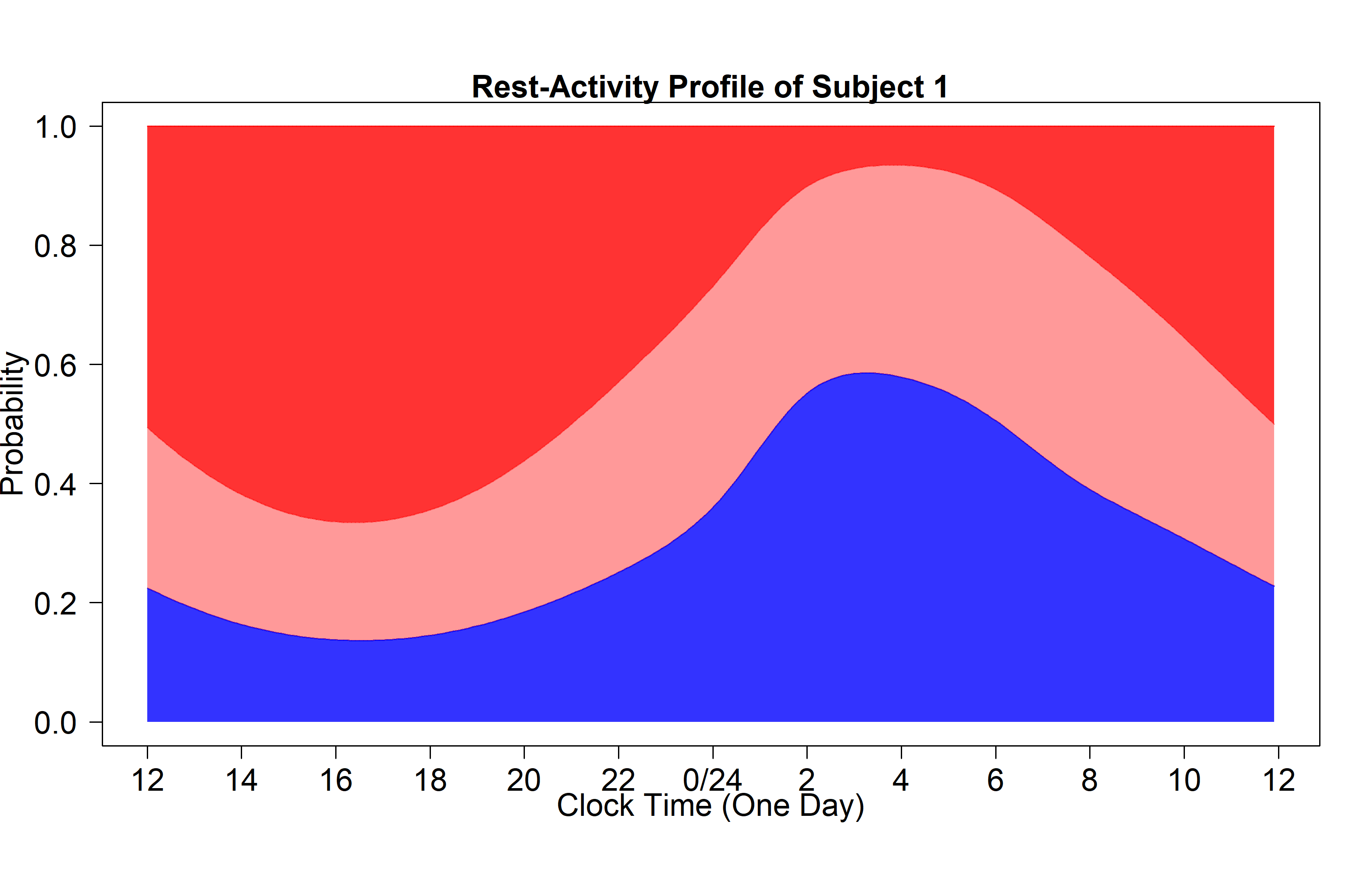} 
    \end{subfigure}
    \hfill
    \begin{subfigure}{0.5\textwidth}
    \includegraphics[width=0.9\linewidth, height=4.4cm]{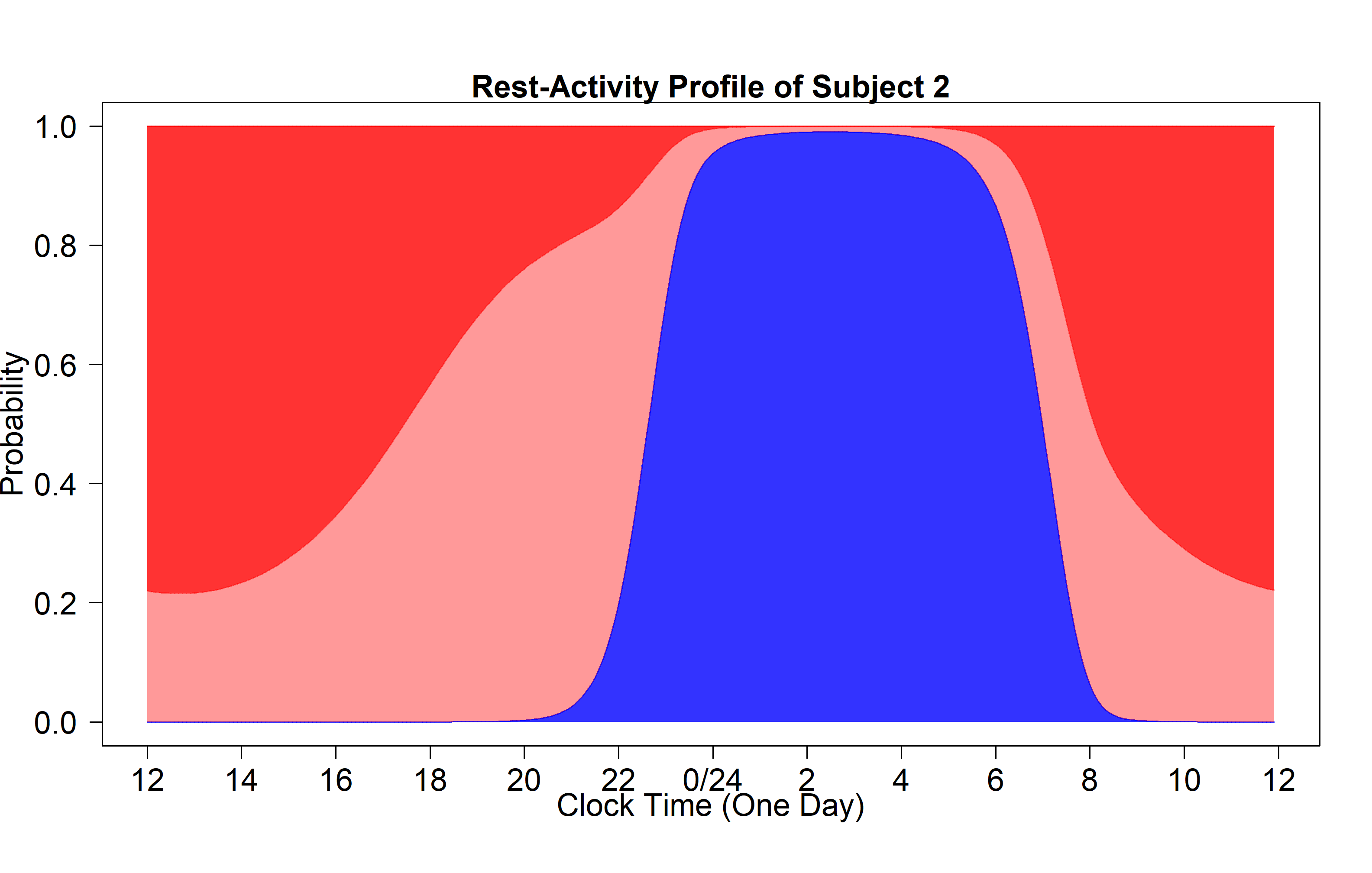}
    \end{subfigure}    

    \caption{Example of the 24-hour rest-activity profile derived from BCHMM estimates for two selected subjects in the 2011-2014 NHANES study. Blue, pink and red areas represent the low-, moderate-, and high-activity state probabilities, respectively.}
    \label{fig:RA_profiles}
\end{figure}

Figure~\ref{fig:RA_profiles} presents two examples of rest-activity profiles, which are essentially three hidden-state probabilities in one 24 hour period from noon (12 pm) to noon. Subject 1 has relatively higher low-activity state probablities throughout the day, but lower during the resting period (0-8 am), observed from the blue shaded area (left panel); while subject 2 has a more concentrated blue shaded area, indicating uninterrupted resting period. Moreover, subject 1 has a lower estimated RI of 0.24, i.e., worse overall rhythmicity, while subject 2 has a higher RI of 0.88, i.e., better overall rhythmicity. These profiles and derived measures (e.g., RI) are consistent with directly observing from the raw actigraphy data (Figure~\ref{fig:raw_act_data})
, but are more visually intuitive and quantitatively informative.

\begin{figure}[htbp]
    \centering
    \includegraphics[width=.9\linewidth]{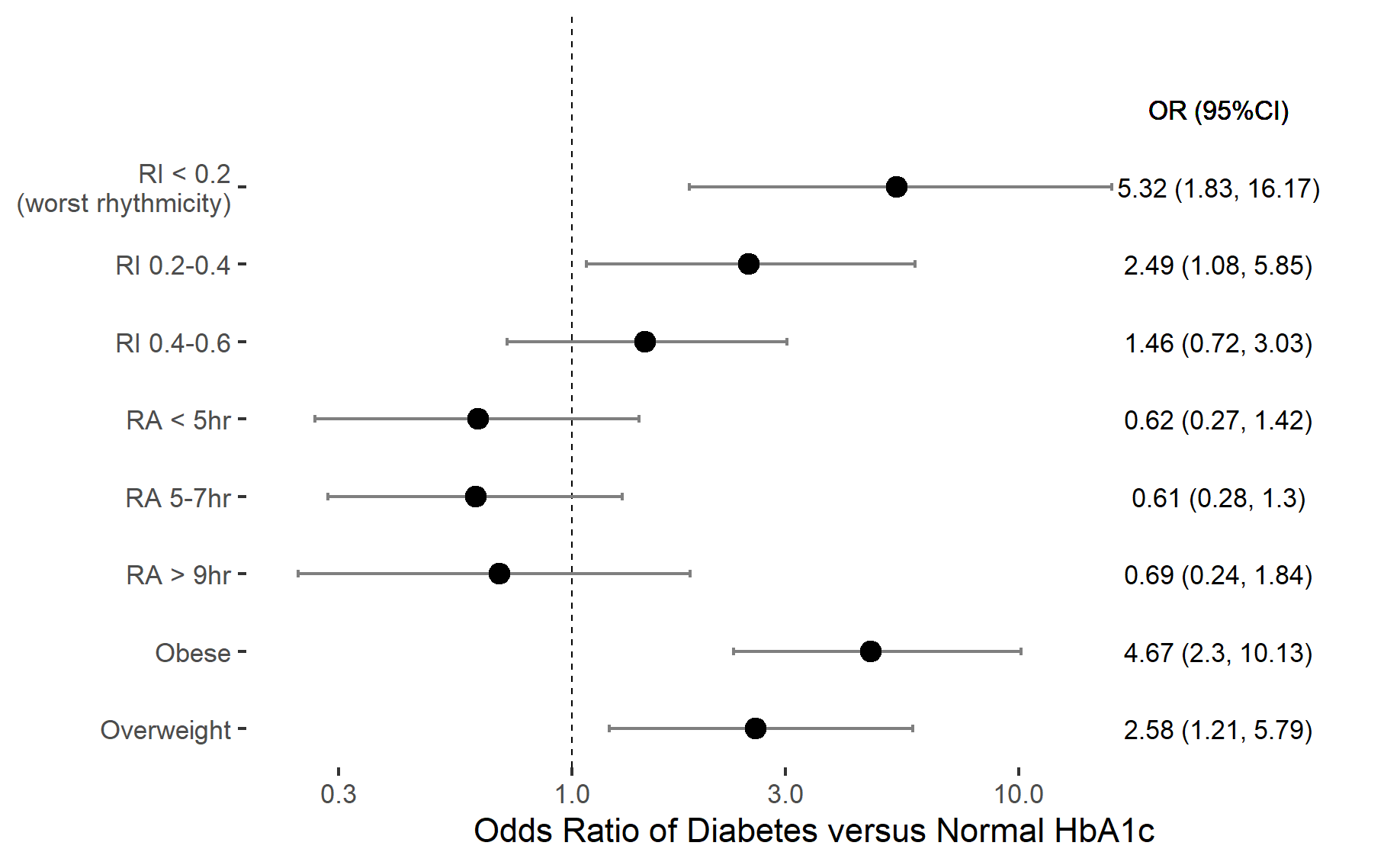} 
    \caption{Association between diabetic status and rest-activity rhythmicity. Results of odds ratios are presented, estimated from the logistic regression model, adjusting for age, gender, weight status characterized by BMI. Increased odds of having diabetes versus normal HbA1c levels is monotonously associated with lower values of RI, i.e., worse overall rhythmicity.}
    \label{fig:odds_ratios_nhanes}

\end{figure}

The derived RAR measures also suggest substantial differences between diabetics and the normal group, where on average, participants with diabetes have significantly lower RI than the normal group (0.440 vs. 0.514, $p$-value = 0.003)(Supplementary Table~\ref{table:nhanes_bchmm_chmm_est_RI} and Figure~\ref{fig:RI_chmm_vs_bchmm_nhanes}). To further quantify the risk of diabetes and the weakened RI, we categorize RI at 0.2 intervals and perform a logistic regression on the diabetic status and RI categories, adjusting for age, gender, weight status, and rest amount (RA) extracted from BCHMM fitting. Weight status is defined by body mass index (BMI, $kg/m^2$): 18.5 and below is classified as underweight, 18.5 - 24.9 as healthy weight, 25.0 - 29.9 as overweight, and 30.0 and above as obese \citep{centers_for_disease_control_and_prevention_all_2022}. Reference levels are normal HbA1c levels, gender of female, normal weight status, best overall rhythmicity of RI $>$ 0.6 and a typical amount of RA between 7 to 9 hours. Figure~\ref{fig:odds_ratios_nhanes} presents the results from the logistic regression. The likelihood of having diabetes, compared to normal HbA1c, progressively increases with weakened rhythmicity of lower RI. The risk of diabetes is much higher among participants with the lowest RI (0.2 and lower), with an estimated odds ratio of 5.32 (95\% CI, [1.83, 16.17]). This result is consistent with existing studies that impaired RAR is strongly associated with diabetic status and impaired glycemic control \citep{xiao_association_2021, sohail_irregular_2015}, which applied the cosinor-based models \citep{cornelissen_cosinor-based_2014} to evaluate the RAR. The advantage of our proposed model lies in its capacity to probabilistically quantify the activity profile. Consequently, it facilitates a more nuanced examination of the differences in 24-hour activity patterns between diabetic and non-diabetic groups.

\section{Discussion}
In this analysis, we introduce a novel Bayesian non-homogeneous Hidden Markov Model that integrates biological circadian rhythms to reflect the rest-activity patterns of each individual. The Bayesian approach effectively resolves the identification or label-switching problem often encountered with frequency-based methods, thereby enhancing statistical inference, as demonstrated in our comprehensive simulation study. We apply our model to 24-hour actigraphy data using 2011-2014 NHANES data to investigate the relationship between rest-activity rhythms and the risk of diabetes. Our proposed methodology reveals nuanced differences in rest-activity profiles associated with diabetes risk that are unattainable with conventional analyses for actigraphy data.

While 24-hour actigraphy data gains considerable popularity in recent years, most analyses concentrate on movement classification. Few studies leverage actigraphy data to comprehend rest-activity rhythms, which are central to our objective. Compared to other methodologies, the Bayesian circadian Hidden Markov model is more complex and demands specialized software for implementation. To ease its adoption, we share our code to enable replication of our analysis. Despite its complexity, the model we propose can better capture the individual rest-activity profiles that other analytical approaches can't, which signifies a significant methodological advantage. Our proposed BCHMM eases the assumption of a certain waveform of rest-activity patterns that may not be applicable to any individual. Meanwhile, our approach offers flexibility to handle missing data and study different rhythmic behaviors. These derived rest-activity measures assess not only certain aspects of the rhythms, as existing approaches commonly provide, but overall rhythmicity, with clinical relevance and epidemiological interpretability. The robustness, flexibility, and interpretability assure promising potential of our model to be applied to broader epidemiological and clinical applications.

Our current model is applied to each individual, yielding personalized rest-activity profiles. This approach is appropriate when sufficient data is available, as in our motivating NHANES dataset, where each individual has ample data to model their rest-activity profile. In addition, we choose to model each person individually due to the inherent heterogeneity of actigraphy data stemming from individual characteristics, disease status, and lifestyle behaviors. However, when working with more limited actigraphy data—either in terms of the number of 24-hour observations or the total number of subjects in the study—a hierarchical Bayesian Hidden Markov model is more suitable, allowing for cross-subject borrowing of rest-activity rhythms. This is our planned research direction moving forward.

\clearpage

\bibliographystyle{rss}
\bibliography{reference_final}

\clearpage

\begin{center}
    \title{{\huge Supplementary Material}}
\end{center}

\setcounter{section}{0}
\setcounter{equation}{0}
\setcounter{figure}{0}
\setcounter{table}{0}
\setcounter{page}{1}
\makeatletter
\renewcommand{\thesection}{S\arabic{section}}
\renewcommand{\theequation}{S\arabic{equation}}
\renewcommand{\thefigure}{S\arabic{figure}}
\renewcommand{\thetable}{S\arabic{table}}
\renewcommand{\bibnumfmt}[1]{[S#1]}
\renewcommand{\citenumfont}[1]{S#1}

\section{Approximation to Hidden Markov Models}

\paragraph{Forward Algorithm}\label{sec:supp_forward}
Inference on parameters in HMMs is obtained by maximum likelihood estimation. With the unobserved hidden states, the marginal likelihood of the observations is required, which can be written in a recursive form as follows,

\begin{equation} \label{marginal likelihood}
\begin{split}
L(\boldsymbol{y}|\Theta) & = P(\boldsymbol{Y}=\boldsymbol{y})\\
    & = \sum_{S_{1},...,S_{T}=1}^{m}P(\boldsymbol{Y}=\boldsymbol{y}, \boldsymbol{S}=\boldsymbol{s})\\
    & = \sum_{S_{1},...,S_{T}=1}^{m}
            \Bigg\{
                P(S_{1}=s_{1})
                \prod_{t=2}^{T}{ P(S_{t}=s_{t}|S_{t-1}=s_{t-1}) }
                \prod_{t=1}^{T}{P(Y_{t}=y_{t}|S_{t}=s_{t})}
            \Bigg\}\\
    & = {
        \delta_{s_1} f(y_{1}|s_1)\gamma_{1,s_{1}s_{2}} f(y_{2}|s_{2})\gamma_{2,s_{2}s_{3}}...\gamma_{T-1,s_{T-1}s_{T}}f(y_{T}|s_{T})
    }\\
    & = \boldsymbol\delta \boldsymbol{P}(y_{1})\Gamma_{1}\boldsymbol{P}(y_{2})\Gamma_{2}\boldsymbol{P}(y_{3})...\Gamma_{T-1}\boldsymbol{P}(y_{T})\boldsymbol1^{'}\\
\end{split}
\end{equation}

\noindent where $\boldsymbol y=(y_1, y_2, ..., y_T)$ is the observation sequence, $\boldsymbol{s}=(s_1, s_2, ..., s_T)$ is the state sequence, and $\boldsymbol{P}(y_{t}) = \Big(f(y_t|S_t=j) \Big)^{'} $.

The maximization of marginal likelihood can be realized using the Forward algorithm. It adopts a so-called predict-update cycle to sequentially determine hidden states that optimize marginal likelihoods. Denote the marginal probability at time $t$ as $\boldsymbol{\alpha_t}=p(s_t|y_{1:t})$ which is a $m\times1$ vector, where $y_{1:t}$ are observations up to the current time point $t$. The \textit{j}-th element of $\boldsymbol{\alpha_t}$ can be calculated as, 
\begin{equation}
    \alpha_t(j) \propto f(y_t|S_t=j)
    \boldsymbol{\gamma}_{t-1, . j}^{'}\boldsymbol{\alpha_{t-1}}\\
\end{equation}
where $f(y_t|S_t=j)$ is the emission probability that represents the \textit{local evidence} at current time point $t$, $\boldsymbol{\gamma}_{t-1, . j}^{'}$ is the transition probabilities from any state to the $j$-th state, and $\boldsymbol{\alpha_{t-1}}$ is the forward probabilities at the previous time point that represents \textit{belif state} \citep{murphy_machine_2012}. In \texttt{depmixS4}, the expectation-maximization (EM) algorithm is employed to estimate the unconstrained models, and direct optimization of a general Newton-Raphson optimizer is used to for models with general linear or non-linear constraints \citep{visser_depmixs4_2010, zucchini_hidden_2016}. In \texttt{Stan}, the forward probabilities are handled in logarithms as
\begin{equation} \label{forward_loglik}
\begin{split}
    \text{log}\big\{\alpha_t(j)\big\}
    & = \text{log}\Big\{
                \sum_{i=1}^{m} p(y_t|s_t=j)\gamma_{ij}\alpha_{t-1}(i)
            \Big\} \\
    & = \text{log}\Big\{
                \sum_{i=1}^{m} exp\big[log(p(y_t|s_t=j)) 
                + \text{log}(\gamma_{ij}) 
                + \text{log}(\alpha_{t-1}(i)) 
                \big]
            \Big\} \\
\end{split}
\end{equation}

\section{Additional Results from the Simulation Study}

\subsection{Circadian Coefficients}

In the simulation, we assume three hidden states (i.e., $m=3$) which give a $3\times 3$ transition probabilities at each time point $t$ as
\begin{equation*}
    \Gamma_t
    = \left( \begin{array}{ccc}
         \gamma_{t,11} & \gamma_{t,12} & \gamma_{t,13}  \\
         \gamma_{t,21} & \gamma_{t,22} & \gamma_{t,23}  \\
         \gamma_{t,31} & \gamma_{t,32} & \gamma_{t,33}  \\
    \end{array} \right) \\
    = \left( \begin{array}{ccc}
         \frac{\text{exp}(\eta_{t,11})}{\sum_{s=1}^{m}\text{exp}(\eta_{t,1s})} &  \frac{\text{exp}(\eta_{t,12})}{\sum_{s=1}^{m}\text{exp}(\eta_{t,1s})} &  \frac{\text{exp}(\eta_{t,13})}{\sum_{s=1}^{m}\text{exp}(\eta_{t,1s})}  \\
         \frac{\text{exp}(\eta_{t,21})}{\sum_{s=1}^{m}\text{exp}(\eta_{t,2s})} &  \frac{\text{exp}(\eta_{t,22})}{\sum_{s=1}^{m}\text{exp}(\eta_{t,2s})} &         \frac{\text{exp}(\eta_{t,23})}{\sum_{s=1}^{m}\text{exp}(\eta_{t,2s})} \\
         \frac{\text{exp}(\eta_{t,31})}{\sum_{s=1}^{m}\text{exp}(\eta_{t,3s})} & 
         \frac{\text{exp}(\eta_{t,32})}{\sum_{s=1}^{m}\text{exp}(\eta_{t,3s})} &  \frac{\text{exp}(\eta_{t,33})}{\sum_{s=1}^{m}\text{exp}(\eta_{t,3s})}  \\
    \end{array} \right) 
\end{equation*}

We assume one pair of circadian oscillators of $\cos(\frac{2\pi t}{24})$ and $\sin(\frac{2\pi t}{24})$, denoted as $\cos_t$ and $\sin_t$ to ease the notation. Hence, $\eta_{t, ij}, i \neq j$ is a linear combination of circadian oscillators at time $t$, and circadian coefficients of $\beta_{ij}$ that explicitly associate with the transition from state $i$ to state $j$, which can be expressed as
\begin{equation*}
    \left( \begin{array}{ccc}
         \eta_{t,i1} & \eta_{t,i2} & \eta_{t,i3}  \\
    \end{array} \right)
    = \left( \begin{array}{ccc}
         1 & \cos_t & \sin_t  \\
    \end{array} \right)
    \left( \begin{array}{ccc}
         \beta_{0,i1} & \beta_{0,i2} & \beta_{0,i3} \\
         \beta_{1,i1} & \beta_{1,i2} & \beta_{1,i3} \\
         \beta_{2,i1} & \beta_{2,i2} & \beta_{2,i3} \\
    \end{array} \right)
\end{equation*}

%

We impose $\boldsymbol{\beta_{ii}}=\boldsymbol{0}$ to ensure parameter identifiability of $\eta_{t,ij}$ as stated in the main text Section~\ref{sec:method}. This leads to assigning values to a set of eighteen coefficients (3 transit-out states $\times$ 2 independent transit-in states $\times$ 3 coefficients for each pair of transitions) for all $\boldsymbol{\beta_{ij}}, i \neq j$. The true values of the circadian coefficients used in the simulation study are listed in Table~\ref{table:simu_true_coef}. These values are informed by the 2011-2014 NHANES actigraphy data that served as our motivating example. The resulting time-varying state probabilities conform to the patterns observed in the real data. Specifically, the probabilities in low-activity (state 1) should be higher during the nighttime and lower during the daytime, and the probabilities in the high-activity (state 3) should be higher during the daytime and lower during the nighttime. These state probabilities are illustrated in Figure~\ref{fig:simulation_est_sp_bias} (a), (c) and (e) as the black dashed lines.


\begin{table}[htbp]
\centering
\caption{True values used for the circadian coefficients in the simulation study.}
\label{table:simu_true_coef}
\begin{tabular}{ c|ccc}\hline
      Hidden State $j$ & low-activity & moderate-activity & high-activity \\
       & $j$=1 & $j$=2 & $j$=3       \\
     \hline
     $\beta_{0, 1j}$  & 0 & -1.89 & -7.27     \\
     $\beta_{1, 1j}$  & 0 & 0.04 & -0.08      \\
     $\beta_{2, 1j}$  & 0 & 0.17 & 3.40       \\
     $\beta_{0, 2j}$  & -4.13 & 0 & -2.78     \\
     $\beta_{1, 2j}$  & 0.11 & 0 & 0.25       \\
     $\beta_{2, 2j}$  & -2.96 & 0 & 0.85      \\
     $\beta_{0, 3j}$  & -8.42 & -2.86 & 0     \\
     $\beta_{1, 3j}$  & -1.07 & 0.04 & 0      \\
     $\beta_{2, 3j}$  & -2.59 & -0.94 & 0     \\
     \hline
\end{tabular}
\end{table}

\subsection{Additional Results from the Proposed BCHMM in the Simulation Study}
We use the posterior median for all parameters from BCHMM and model performance evaluation metrics described in main text Equation~\eqref{eq:simualtion_metrics} and \eqref{eq:acc_sp_bias}. The results, presented in Table~\ref{table:bchmm_results}, show that BCHMM has desirable performance in correctly estimating the distributional parameters and most circadian coefficients. This is evidenced by the smaller values of MAB, MMSE, and MSD. Furthermore, the MCRs for the mean and variance parameters are predominantly above 95\% across all simulation scenarios.

\begin{table}[htbp]\centering
\caption{Evaluation of the parameter estimation from BCHMM.}
\label{table:bchmm_results}

\begin{tabular}[t]{c p{.8cm}p{.8cm}p{.8cm}p{.8cm}p{.8cm}p{.8cm}p{.8cm}p{.8cm}p{.8cm}p{.8cm}p{.8cm}p{.8cm}}
\hline
 & \multicolumn{4}{c}{Scenario 1} & \multicolumn{4}{c}{Scenario 2} & \multicolumn{4}{c}{Scenario 3}\\
Parameter & MAB$^\dag$ & MMSE$^\ddag$ & MSD$^{\S}$ & MCR$^{\S\S}$ & MAB & MMSE & MSD & MCR & MAB & MMSE & MSD & MCR\\
\hline
$\mu_1$ & 0.033 & 0.002 & 0.042 & 95\% & 0.056 & 0.005 & 0.062 & 91\% & 0.040 & 0.002 & 0.048 & 98\%\\
 
$\mu_2$ & 0.026 & 0.001 & 0.032 & 93\% & 0.038 & 0.002 & 0.047 & 97\% & 0.048 & 0.004 & 0.063 & 96\%\\
 
$\mu_3$ & 0.021 & 0.001 & 0.028 & 95\% & 0.032 & 0.002 & 0.039 & 96\% & 0.044 & 0.003 & 0.059 & 95\%\\
 
$\sigma^2_1$ & 0.033 & 0.002 & 0.044 & 93\% & 0.075 & 0.008 & 0.092 & 97\% & 0.044 & 0.003 & 0.055 & 95\%\\
 
$\sigma^2_2$ & 0.027 & 0.001 & 0.033 & 92\% & 0.055 & 0.005 & 0.069 & 95\% & 0.106 & 0.016 & 0.125 & 96\%\\
 
$\sigma^2_3$ & 0.018 & 0.000 & 0.028 & 99\% & 0.048 & 0.004 & 0.056 & 92\% & 0.098 & 0.016 & 0.128 & 94\%\\
 
$\beta_{0,12}$ & 0.479 & 0.449 & 0.632 & 97\% & 0.482 & 0.447 & 0.632 & 94\% & 0.505 & 0.414 & 0.676 & 94\%\\
 
$\beta_{1,12}$ & 0.301 & 0.135 & 0.360 & 95\% & 0.289 & 0.153 & 0.374 & 94\% & 0.324 & 0.195 & 0.387 & 92\%\\
 
$\beta_{2,12}$ & 0.606 & 0.690 & 0.761 & 97\% & 0.601 & 0.691 & 0.762 & 93\% & 0.601 & 0.579 & 0.808 & 97\%\\
 
$\beta_{0,13}$ & 5.217 & 31.088 & 5.723 & 98\% & 5.425 & 33.331 & 5.666 & 100\% & 4.842 & 27.126 & 5.711 & 100\%\\
 
$\beta_{1,13}$ & 1.117 & 2.101 & 5.912 & 100\% & 0.936 & 1.272 & 5.901 & 100\% & 1.251 & 2.682 & 5.862 & 100\%\\
 
$\beta_{2,13}$ & 1.597 & 4.024 & 7.337 & 100\% & 1.361 & 4.297 & 7.299 & 99\% & 1.877 & 5.682 & 7.425 & 100\%\\
 
$\beta_{0,21}$ & 0.479 & 0.365 & 0.554 & 95\% & 0.489 & 0.419 & 0.582 & 92\% & 0.441 & 0.348 & 0.601 & 95\%\\
 
$\beta_{1,21}$ & 0.295 & 0.133 & 0.335 & 94\% & 0.291 & 0.132 & 0.349 & 89\% & 0.261 & 0.111 & 0.354 & 96\%\\
 
$\beta_{2,21}$ & 0.568 & 0.523 & 0.678 & 95\% & 0.585 & 0.572 & 0.715 & 94\% & 0.524 & 0.488 & 0.732 & 95\%\\
 
$\beta_{0,23}$ & 0.173 & 0.053 & 0.224 & 94\% & 0.191 & 0.068 & 0.225 & 91\% & 0.176 & 0.052 & 0.229 & 96\%\\
 
$\beta_{1,23}$ & 0.214 & 0.078 & 0.275 & 93\% & 0.241 & 0.087 & 0.278 & 92\% & 0.248 & 0.094 & 0.284 & 93\%\\
 
$\beta_{2,23}$ & 0.323 & 0.187 & 0.353 & 92\% & 0.291 & 0.122 & 0.351 & 94\% & 0.246 & 0.101 & 0.356 & 98\%\\
 
$\beta_{0,31}$ & 6.514 & 48.237 & 4.888 & 82\% & 6.875 & 51.165 & 5.063 & 91\% & 6.507 & 46.993 & 5.009 & 91\%\\
 
$\beta_{1,31}$ & 1.158 & 3.087 & 5.254 & 97\% & 0.982 & 2.133 & 5.486 & 100\% & 1.204 & 3.083 & 5.446 & 100\%\\
 
$\beta_{2,31}$ & 1.801 & 3.948 & 5.477 & 100\% & 1.727 & 3.934 & 5.757 & 99\% & 1.712 & 4.056 & 5.593 & 99\%\\
 
$\beta_{0,32}$ & 0.204 & 0.061 & 0.236 & 97\% & 0.204 & 0.067 & 0.236 & 91\% & 0.187 & 0.061 & 0.242 & 91\%\\
 
$\beta_{1,32}$ & 0.220 & 0.077 & 0.296 & 94\% & 0.248 & 0.098 & 0.296 & 93\% & 0.254 & 0.115 & 0.305 & 91\%\\
 
$\beta_{2,32}$ & 0.265 & 0.112 & 0.344 & 94\% & 0.285 & 0.125 & 0.348 & 96\% & 0.250 & 0.108 & 0.348 & 95\%\\
\hline
\multicolumn{13}{l}{Definition of the following evaluation metrics can be found in Equation~\ref{eq:simualtion_metrics}.}\\
\multicolumn{13}{l}{$^\dag$Mean absolute bias.}\\
\multicolumn{13}{l}{$^\ddag$Mean of mean squared error.}\\
\multicolumn{13}{l}{$^\S$Mean of posterior standard deviation.}\\
\multicolumn{13}{l}{$^{\S\S}$Mean of coverage rate.}\\
\hline
\end{tabular}

\end{table}

\subsection{Examples of MCMC Diagnostic Plots}

\begin{figure}[htbp]
    \centering
    \includegraphics[width=1\linewidth]{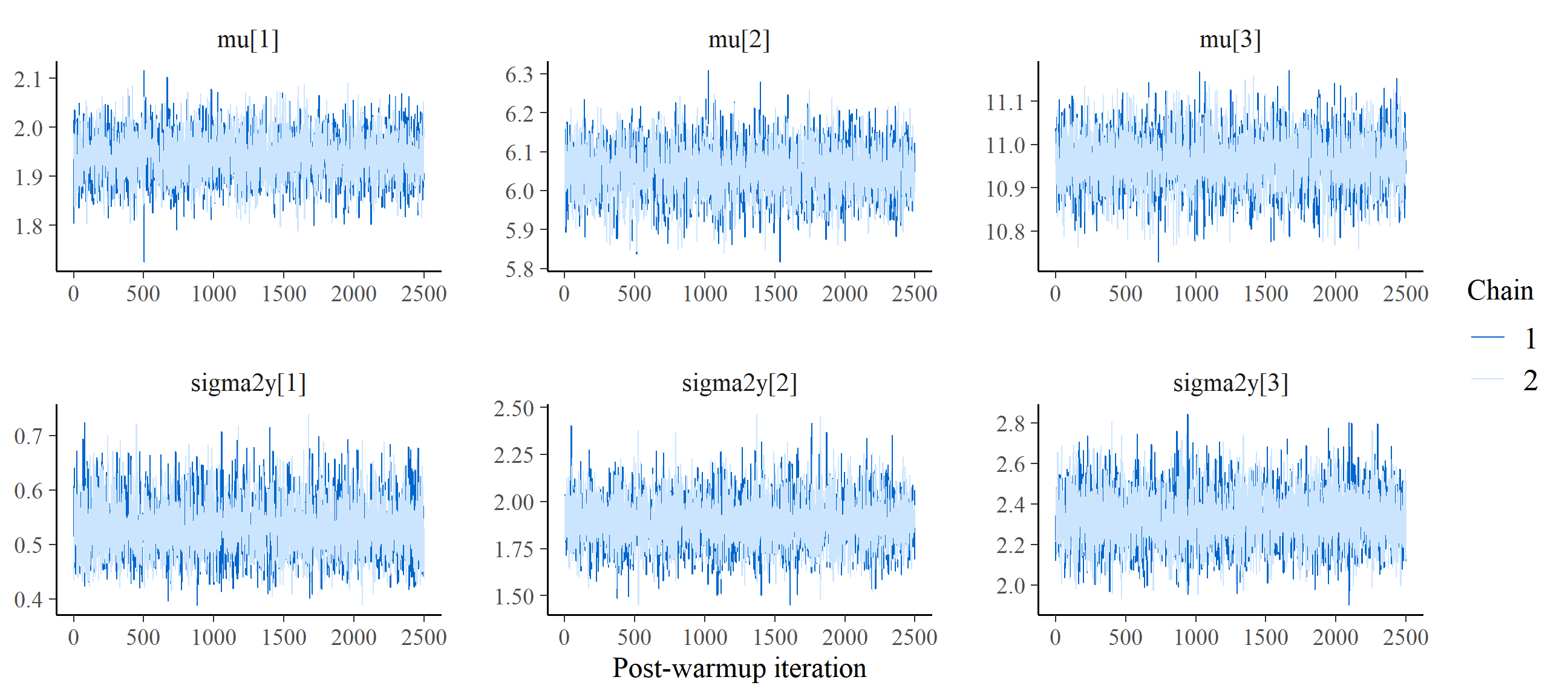} 
    \caption{Example trace plots for means and variances from one replicate under simulation scenario 3.}
    \label{fig:traceplot_means_vars}
\end{figure}

\begin{figure}
    \centering
    \includegraphics[width=1\linewidth]{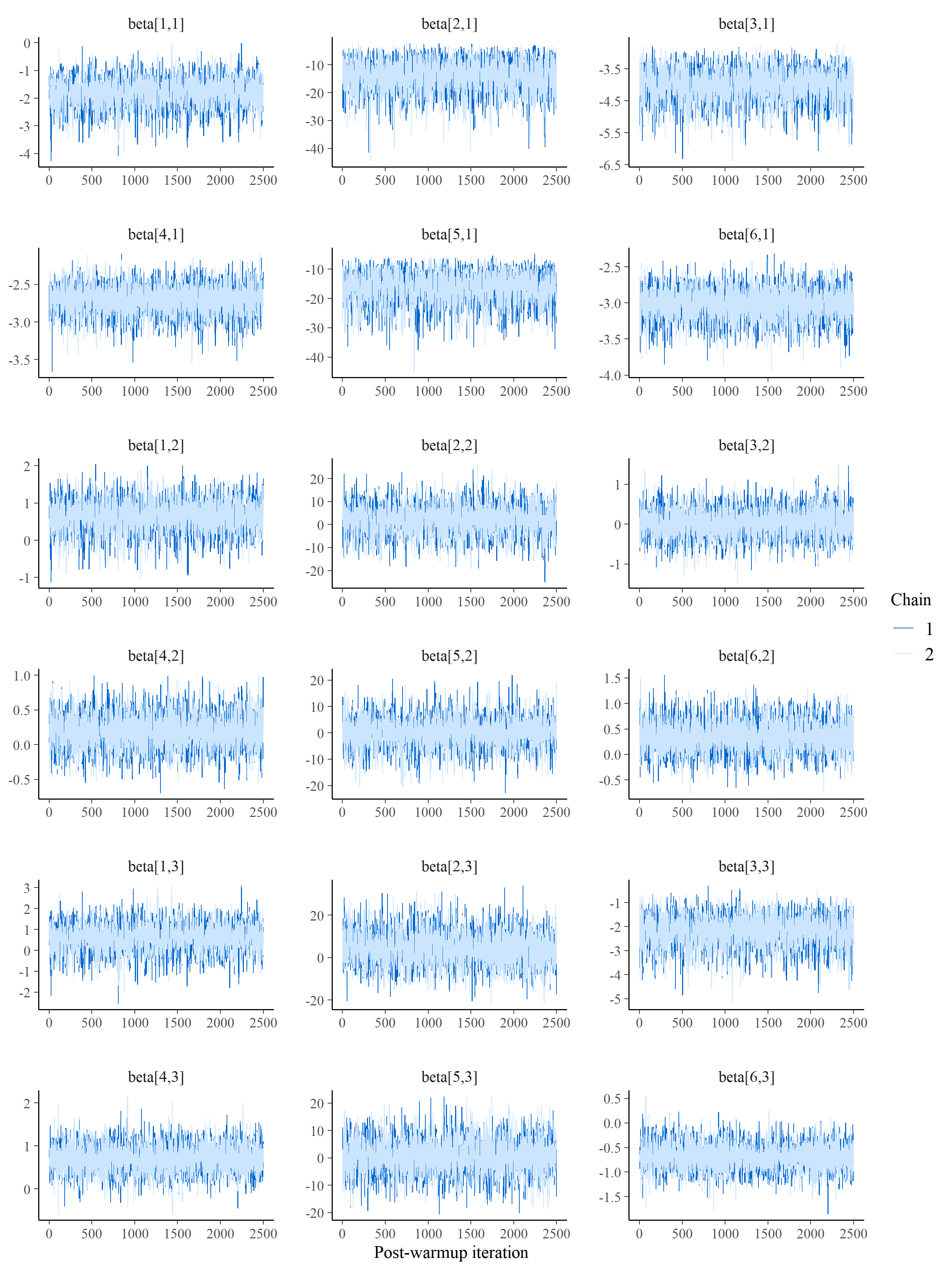} 
    \caption{Example trace plots for circadian coefficients from one replicate under simulation scenario 3.}
    \label{fig:traceplot_betas}
\end{figure}

Figures~\ref{fig:traceplot_means_vars} and \ref{fig:traceplot_betas} present the trace plots of one replicate under simulation scenario 3. The results indicate that two MCMC chains mixed well.


\clearpage

\subsection{Comparison between BCHMM and CHMM on the Emission Distributions}

\begin{figure}[htbp]
    \centering
    \includegraphics[width=.9\linewidth]{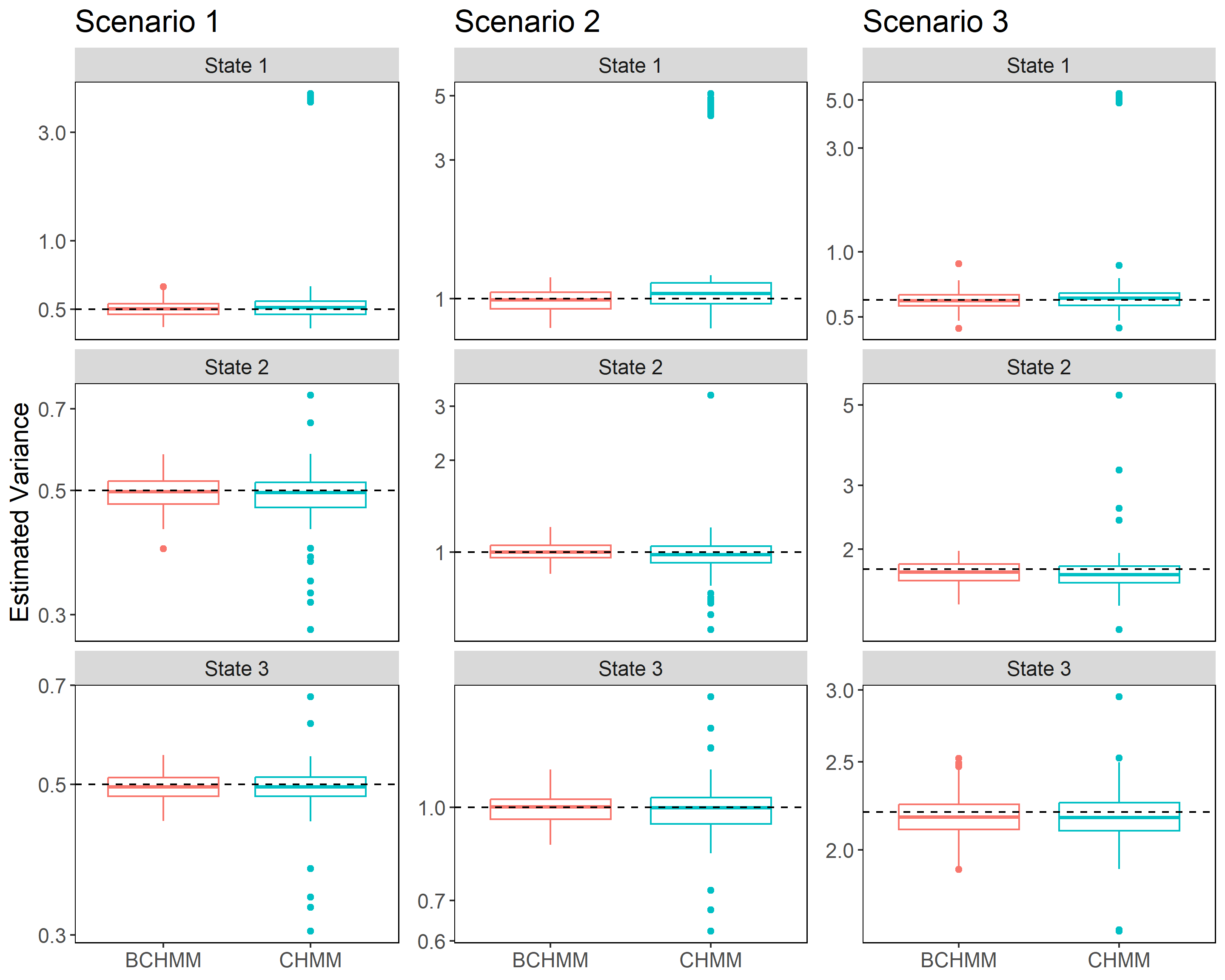} 
    \caption{Comparison of estimated state variances from BCHMM and CHMM. True values are indicated in black dashed lines.}
    \label{fig:var_chmm_vs_bchmm}

\end{figure}

In the main text, we present the estimated means of the hidden states from both BCHMM and CHMM. We provide additional results comparing BCHMM and CHMM in the simulation study. Figure~\ref{fig:var_chmm_vs_bchmm} shows the bias of the estimated state variances (i.e., $\sigma^2_i)$ from these two approaches across the 100 replicates in the simulation study. BCHMM has noticeably more precise estimates than CHMM. The exceptionally high estimated variances from CHMM are likely due to its inability to distinguish low and moderate activity states and over-estimation of the mean value in the low activity state (state 1). 

\subsection{Comparison between BCHMM and CHMM on the Time-varying State Probabilities}

\begin{figure}[htbp]
    \begin{subfigure}{0.5\textwidth}
    \includegraphics[width=0.9\linewidth]{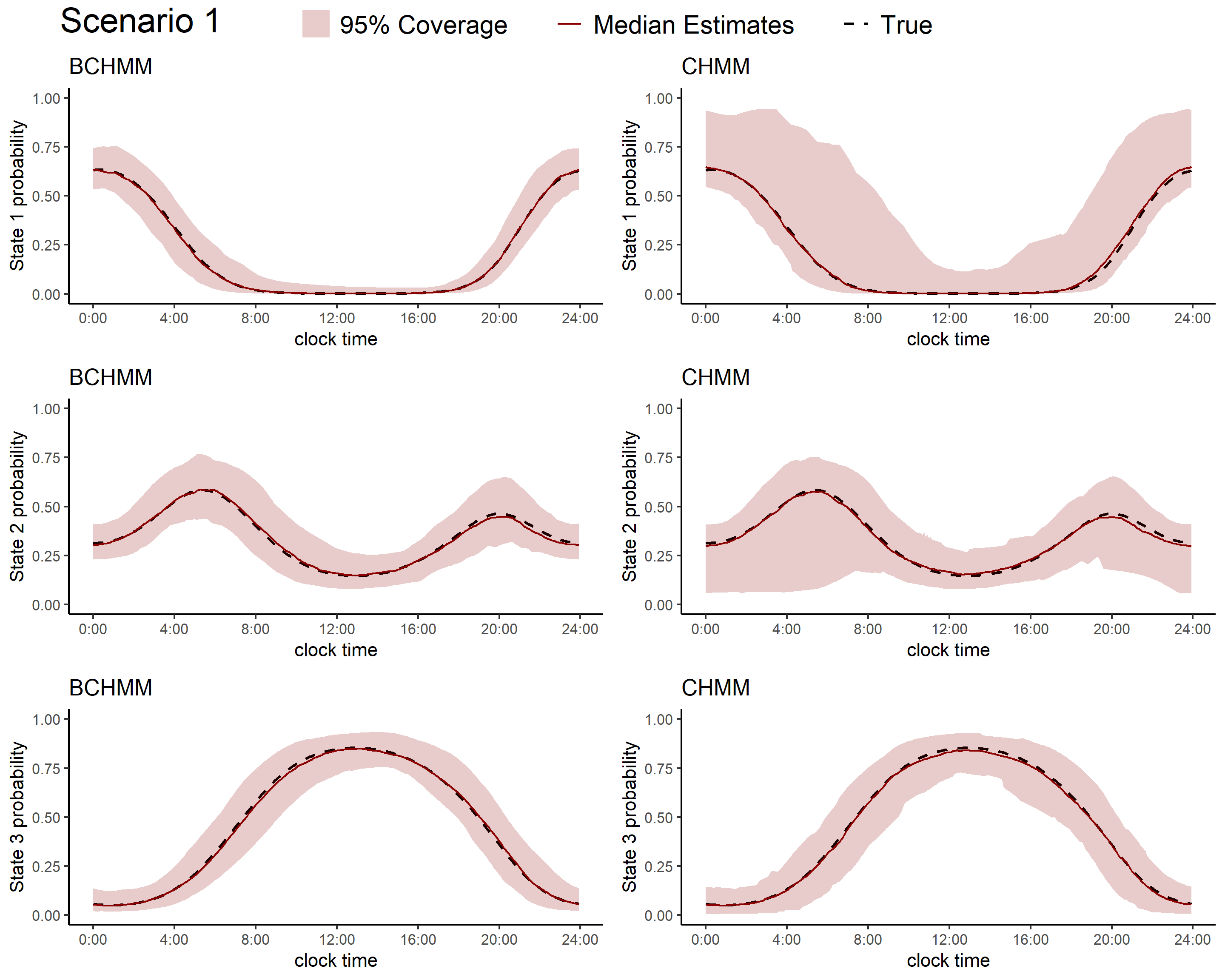} 
    \caption{Scenario 1, Estimated state probabilities}\label{fig:s1_est_sp}
    \end{subfigure}
    \hfill
    \begin{subfigure}{0.5\textwidth}
    \includegraphics[width=0.9\linewidth]{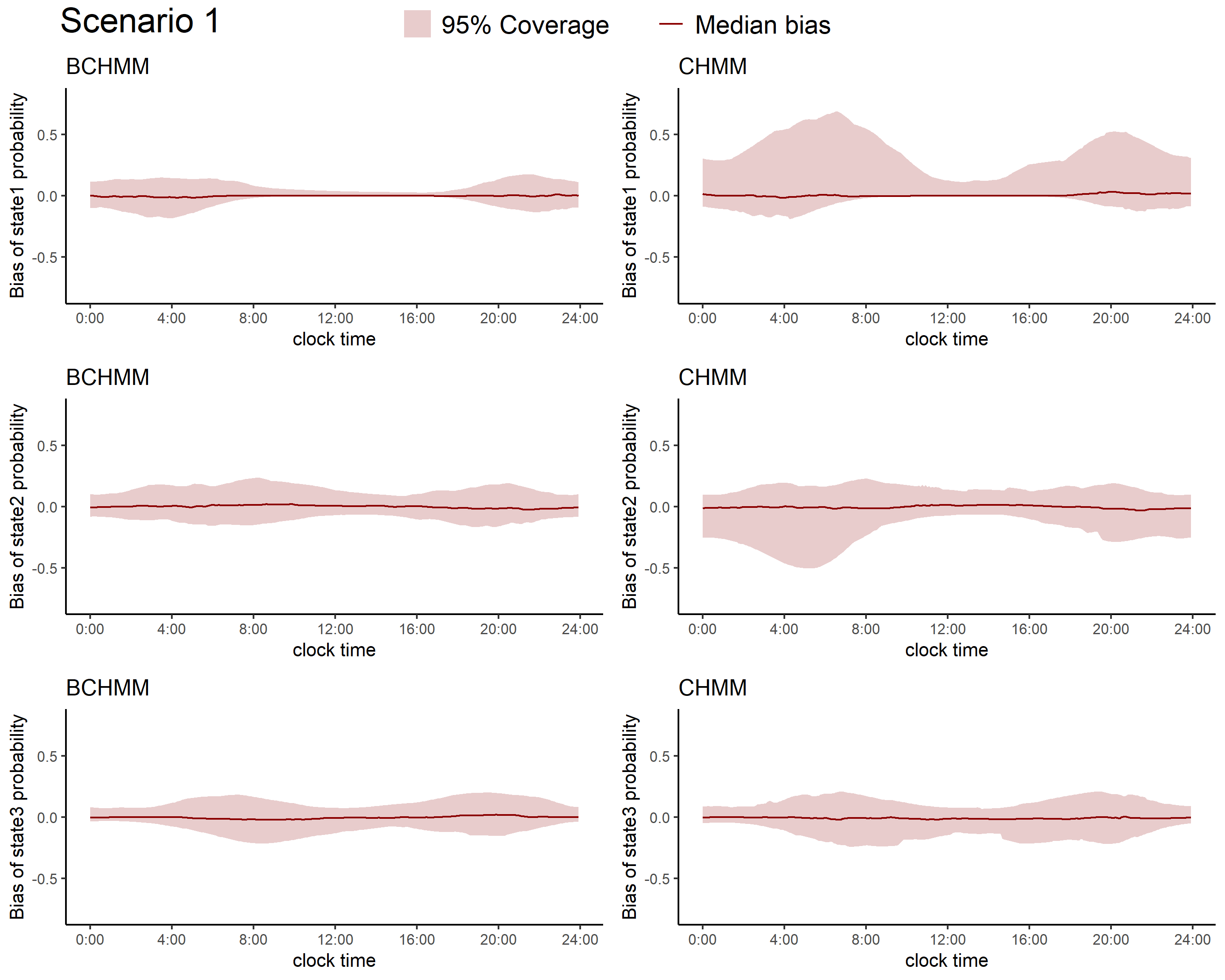}
    \caption{Scenario 1, Bias on estimates versus true}\label{fig:s1_sp_bias}
    \end{subfigure}    
    \hfill
    \begin{subfigure}{0.5\textwidth}
    \includegraphics[width=0.9\linewidth]{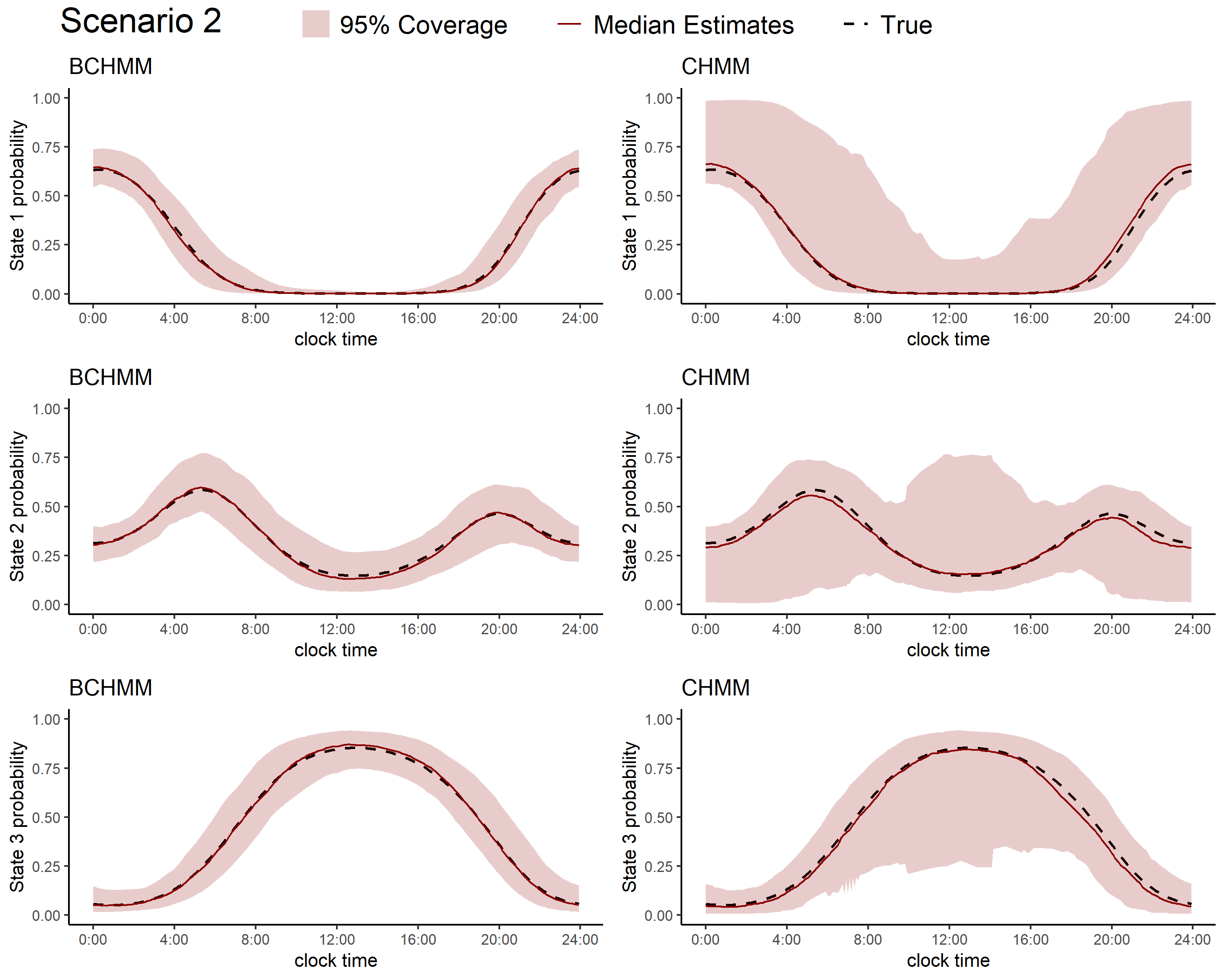} 
    \caption{Scenario 2, Estimated state probabilities}\label{fig:s2_est_sp}
    \end{subfigure}
    \hfill
    \begin{subfigure}{0.5\textwidth}
    \includegraphics[width=0.9\linewidth]{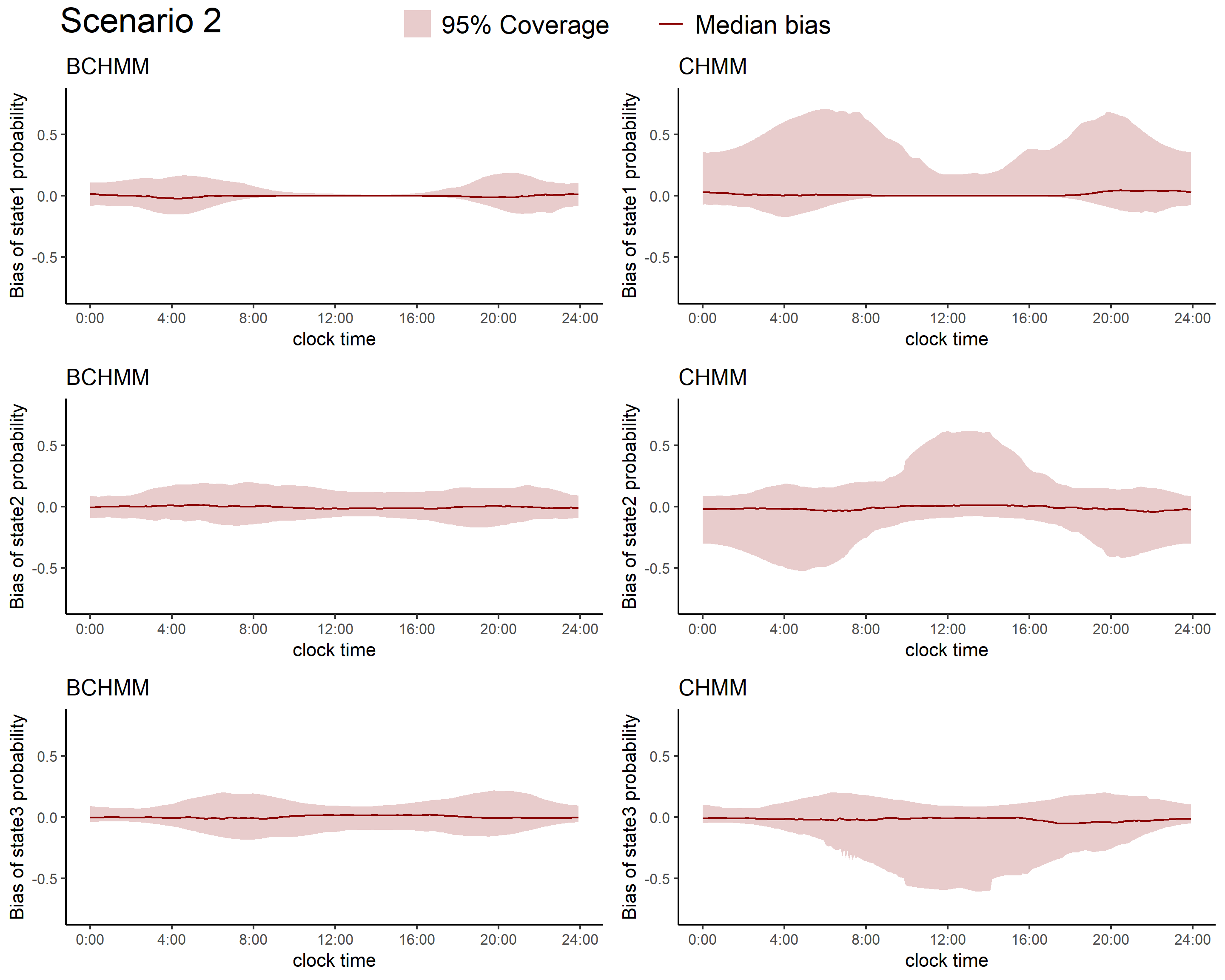}
    \caption{Scenario 2, Bias on estimates versus true}\label{fig:s2_sp_bias}
    \end{subfigure}    
    \hfill
    \begin{subfigure}{0.5\textwidth}
    \includegraphics[width=0.9\linewidth]{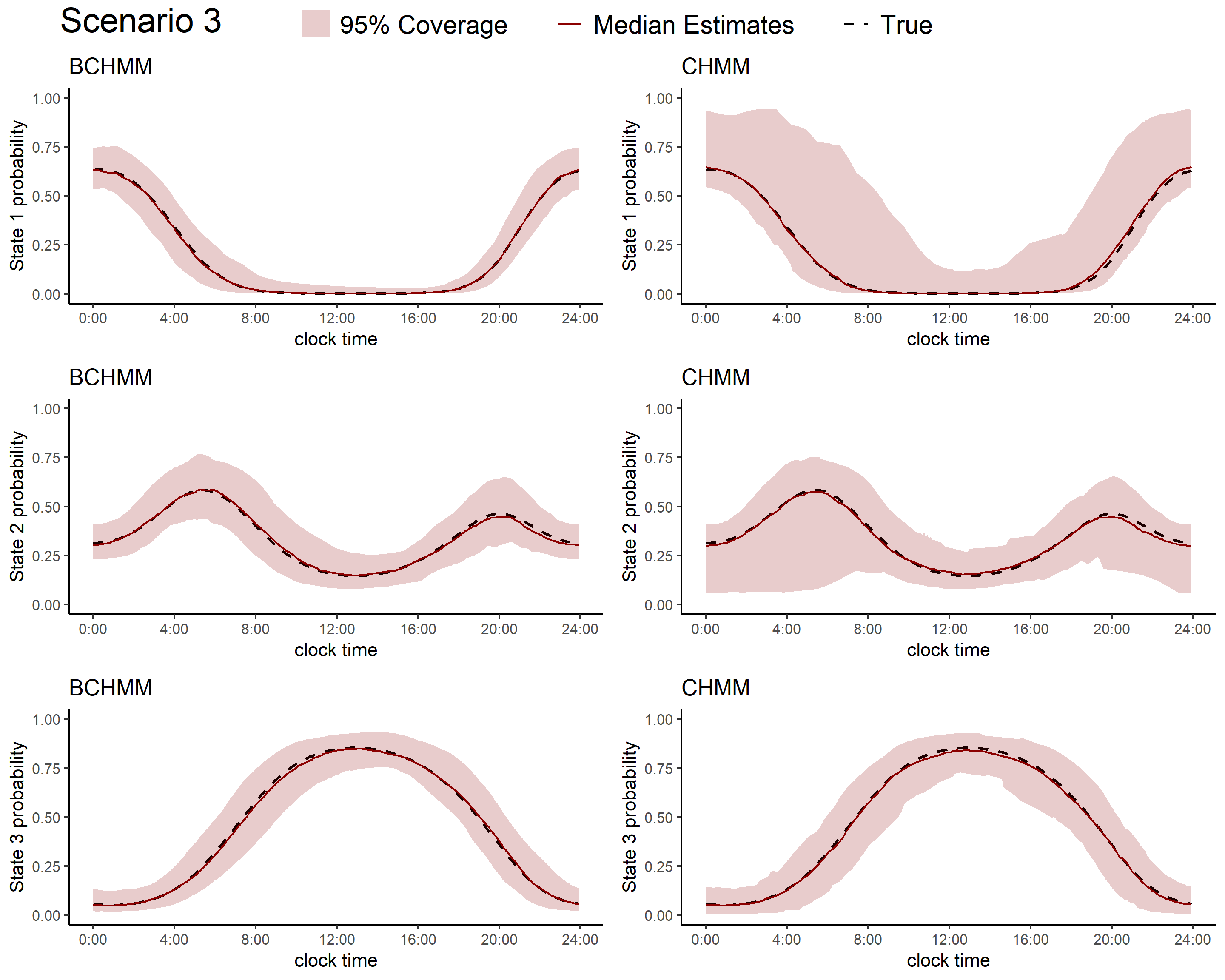} 
    \caption{Scenario 3, Estimated state probabilities}\label{fig:s3_est_sp}
    \end{subfigure}
    \hfill
    \begin{subfigure}{0.5\textwidth}
    \includegraphics[width=0.9\linewidth]{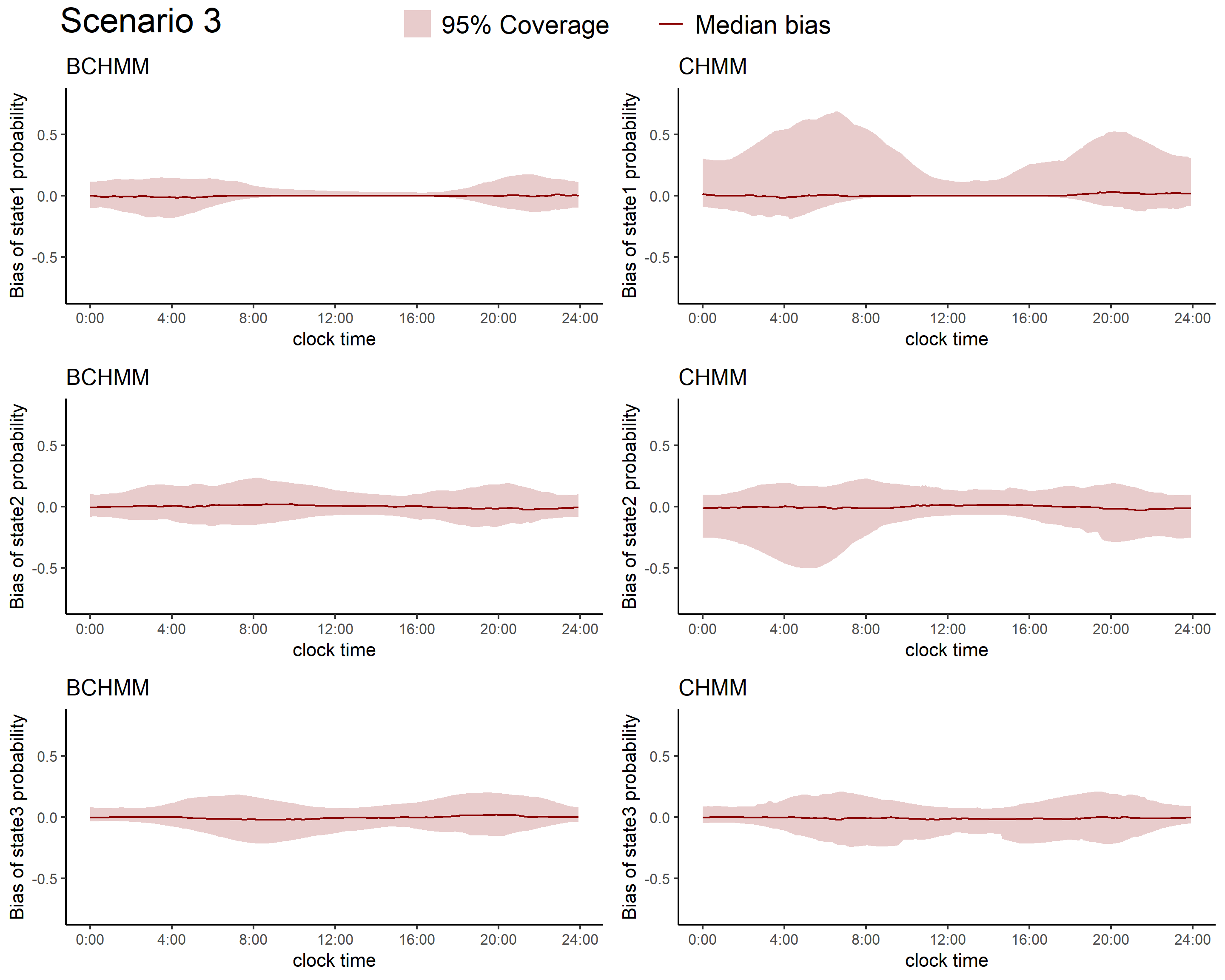}
    \caption{Scenario 3, Bias on estimates versus true}\label{fig:s3_sp_bias}
    \end{subfigure}    
    \caption{Estimated time-varying state probabilities $\hat{P}(S_t)$ (Equation~\eqref{eq:sp_bias}) and bias (calculated as the estimated value minus true, i.e., $\hat{P}(S_t) - P_{True}(S_t)$) from the simulation study. Figures on the left two columns present true state probabilities (black dashed line), median estimates (red solid line) and 95\% coverage band (lower 2.5\% and upper 97.5\% estimates) across 100 replicates. Figures on the right two columns present median bias (red solid line) and 95\% coverage band across 100 replicates. BCHMM generates less biased estimates (more aligned red solid line and black dashed line on the left panels) and more precise estimates with narrower 95\% coverage.}
    \label{fig:simulation_est_sp_bias}

\end{figure}

\begin{figure}[htbp]
    
    \includegraphics[width=1\linewidth]{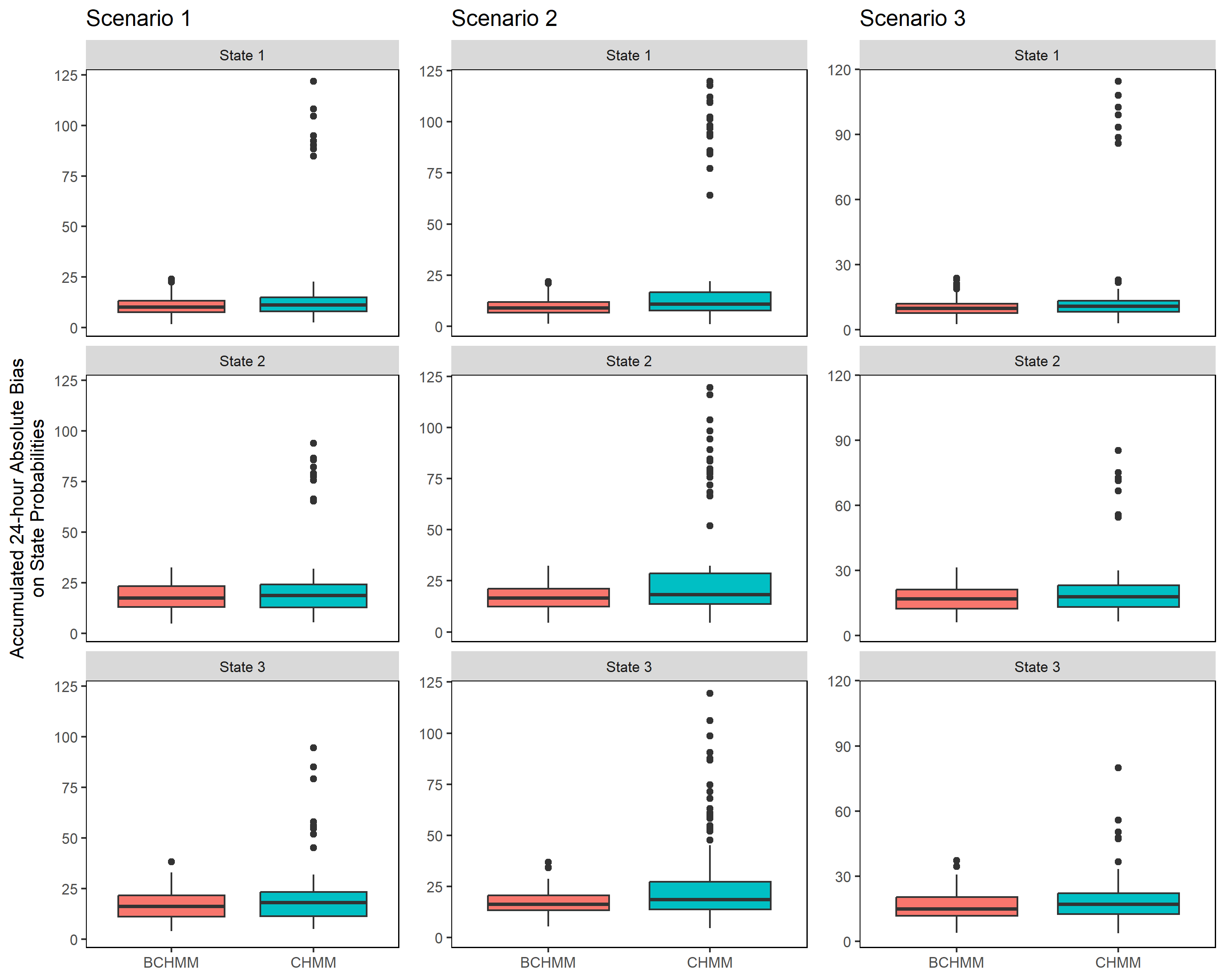} 
    \caption{24-hour accumulated absolute bias.}
    \label{fig:simulation_sp_acc_bias}

\end{figure}

Figure~\ref{fig:simulation_est_sp_bias} presents the estimated time-varying state probabilities for a 24-h period, compared to the true values (black dashed lines). As shown in Figures~\ref{fig:simulation_est_sp_bias} (a, c, e), BCHMM remains high accuracy and precision in recovering the true values, with median estimates (red solid lines) closely following the true values (black dashed lines) and narrower 95\% coverage band (red shaded area). On the other hand, CHMM substantially overestimates the state means, particularly for state 1, as seen from a wider 95\% coverage band above true values (Figures~\ref{fig:simulation_est_sp_bias} (a, c, e)) and the bias from estimates to true state probabilities significantly above zero (Figures~\ref{fig:simulation_est_sp_bias} (b, d, f)). We also evaluate accumulated absolute bias for state probabilities presented in Figure~\ref{fig:simulation_sp_acc_bias}. CHMM, in comparison to BCHMM, tends to demonstrate a higher average accumulated absolute bias and more extreme values.

\section{Additional Results Comparing BCHMM and CHMM on the NHANES Data}

\begin{figure}[htbp]
    \centering
    \includegraphics[width=.9\linewidth]{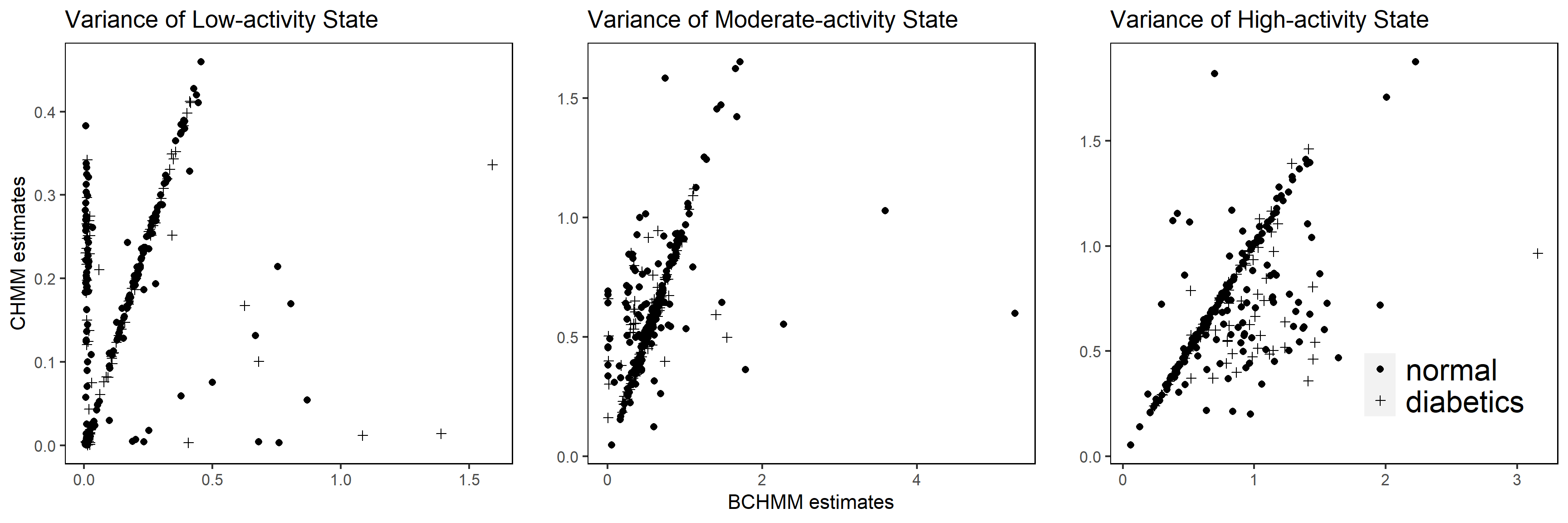} 
    \caption{Comparison of estimated state variances from BCHMM and CHMM on NHANES participants. BCHMM estimates are on $X$ axis while CHMM estimates on $Y$ axis. The solid dots represent participants with diabetes while plus symbol represents the group with normal HbA1c.}
    \label{fig:var_chmm_vs_bchmm_nhanes}

\end{figure}

\begin{figure}[htbp]
    \centering
    \includegraphics[width=.7\linewidth]{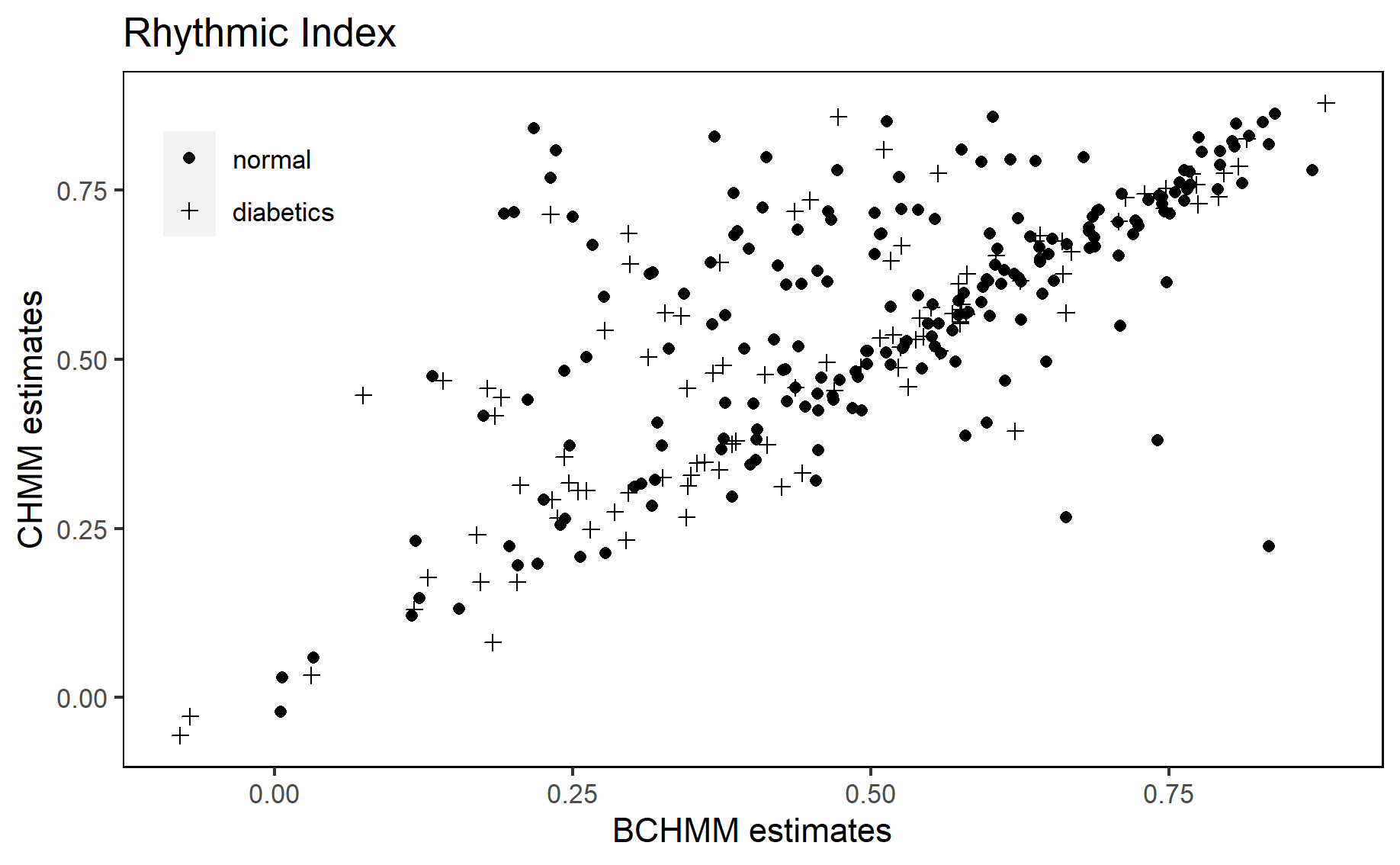} 
    \caption{Comparison of estimated rhythmic index from BCHMM and CHMM on NHANES participants. BCHMM estimates are on $X$ axis while CHMM estimates on $Y$ axis. The solid dots represent participants with diabetes while plus symbol represents the group with normal HbA1c.}
    \label{fig:RI_chmm_vs_bchmm_nhanes}

\end{figure}

We apply both BCHMM and CHMM to the analytical samples derived from 2011-2014 NHANES actigraphy dataset, and compare the estimates from both methods. In the main text, we discuss that CHMM tends to overestimate state means, particularly for the low-activity state (i.e., state 1). This overestimation leads to higher variance estimation of the low-activity state in CHMM compared to BCHMM, as seen in the leftmost panel in Figure~\ref{fig:var_chmm_vs_bchmm_nhanes}. Comparison of model parameters and rest-activity rhythm measures \citep{huang_hidden_2018} by the normal and diabetic groups, are presented in Table~\ref{table:nhanes_bchmm_chmm_est_RI} and Figure~\ref{fig:RI_chmm_vs_bchmm_nhanes}. Both models show that compared to the normal HbA1C group, the diabetic group exhibits significantly lower mean values in the high-activity states (i.e., $\mu_3$) and rhythm indices (RIs), indicating that individuals with diabetes are not only less active but also exhibit worse rest-activity rhythms. However, CHMM also identifies a significant difference in moderate-activity means between the two groups, which is not found in BCHMM. We also perform the logistic regression model on the diabetic status and rest-activity rhythm measures as discussed in the main text Section~\ref{sec:application} with CHMM estimates. While the overall conclusion that worsened overall rhythmicity, indicated by lower RI values, is associated with a higher likelihood of diabetes risk remains similar, the correlation is less pronounced by the CHMM. The odds ratios of diabetes risk associated with weakened RI are 1.46, 2.49, 5.32 with BCHMM, versus 1.85, 3.36, and 3.86 with CHMM. In conclusion, BCHMM offers more precise estimates, potentially highlighting and clarifying the associations between rest-activity rhythm measures and relevant health outcomes more effectively.


\begin{table}[ht]
\centering
\caption{Comparison of estimated parameters from BCHMM and CHMM on NHANES sample.}
\label{table:nhanes_bchmm_chmm_est_RI}

\begin{tabular}[t]{l lll lll}
\hline
 & \multicolumn{3}{c}{BCHMM} & \multicolumn{3}{c}{CHMM} \\
 & Diabetics & Normal & $P$-value$^\dag$ & Diabetics & Normal & $P$-value$^\dag$\\
 \hline
\multicolumn{4}{l}{\textbf{State Mean, Mean (SD)}}\\
{ }Low-activity & 0.270 (0.215) & 0.294 (0.225) & 0.387
 & 0.369 (0.202) & 0.372 (0.210) & 0.906 
 \\
{ }Moderate-activity & 1.73 (0.860) & 1.90 (0.832) & 0.104
 & 1.97 (0.729) & 2.16 (0.707) & 0.0301 \\
{ }High-activity & 3.80 (0.701) & 4.05 (0.658) & 0.004 
 & 3.93 (0.662) & 4.19 (0.627) & 0.001
 \\
\textbf{RI, Mean (SD)} & 0.440 (0.209) & 0.514 (0.190) & 0.003 
 & 0.496 (0.204) & 0.575 (0.185) & 0.001 
 \\
\multicolumn{4}{l}{\textbf{RI Category, N(\%)}}\\
 {    }  $<$ 0.2 & 13 (13.0\%) & 11 (5.5\%) & 0.006 
 & 8 (8.0\%) & 8 (4.0\%) & 0.004 
 \\
 {    }  0.2 - 0.4 & 32 (32.0\%) & 42 (21.0\%) & 
 & 25 (25.0\%) & 26 (13.0\%) & \\
 {    }  0.4 - 0.6 & 33 (33.0\%) & 76 (38.0\%) & 
 & 34 (34.0\%) & 62 (31.0\%) & \\
 {    }  $>$ 0.6 & 22 (22.0\%) & 71 (35.5\%) & 
 & 33 (33.0\%) & 104 (52.0\%) & \\
\textbf{RA, Mean (SD)} & 6.10 (2.21) & 6.28 (2.04) & 0.482
 & 6.89 (1.94) & 6.98 (1.84) & 0.703
 \\
\multicolumn{4}{l}{\textbf{RA Category, N(\%)}}\\
 {    } $<$ 5 hr & 38 (38.0\%) & 62 (31.0\%) & 0.609 &
 22 (22.0\%) & 29 (14.5\%) & 0.373 
 \\
 {    } 5 - 7 hr & 31 (31.0\%) & 73 (36.5\%) & 
 & 27 (27.0\%) & 67 (33.5\%) &\\
 {    } 7 - 9 hr & 22 (22.0\%) & 43 (21.5\%) & 
 & 38 (38.0\%) & 78 (39.0\%) &\\
 {    } $>$ 9 hr & 9 (9.0\%) & 22 (11.0\%) & 
 & 13 (13.0\%) & 26 (13.0\%) &\\
\hline
\multicolumn{7}{p{16cm}}{$^\dag$Two sample $t$-tests or $\chi^2$ tests are used to test the difference on estimated values of parameters between diabetes and normal HbA1c groups.}\\
\end{tabular}

\end{table}

\begin{figure}[htbp]
    \centering
    \includegraphics[width=.9\linewidth]{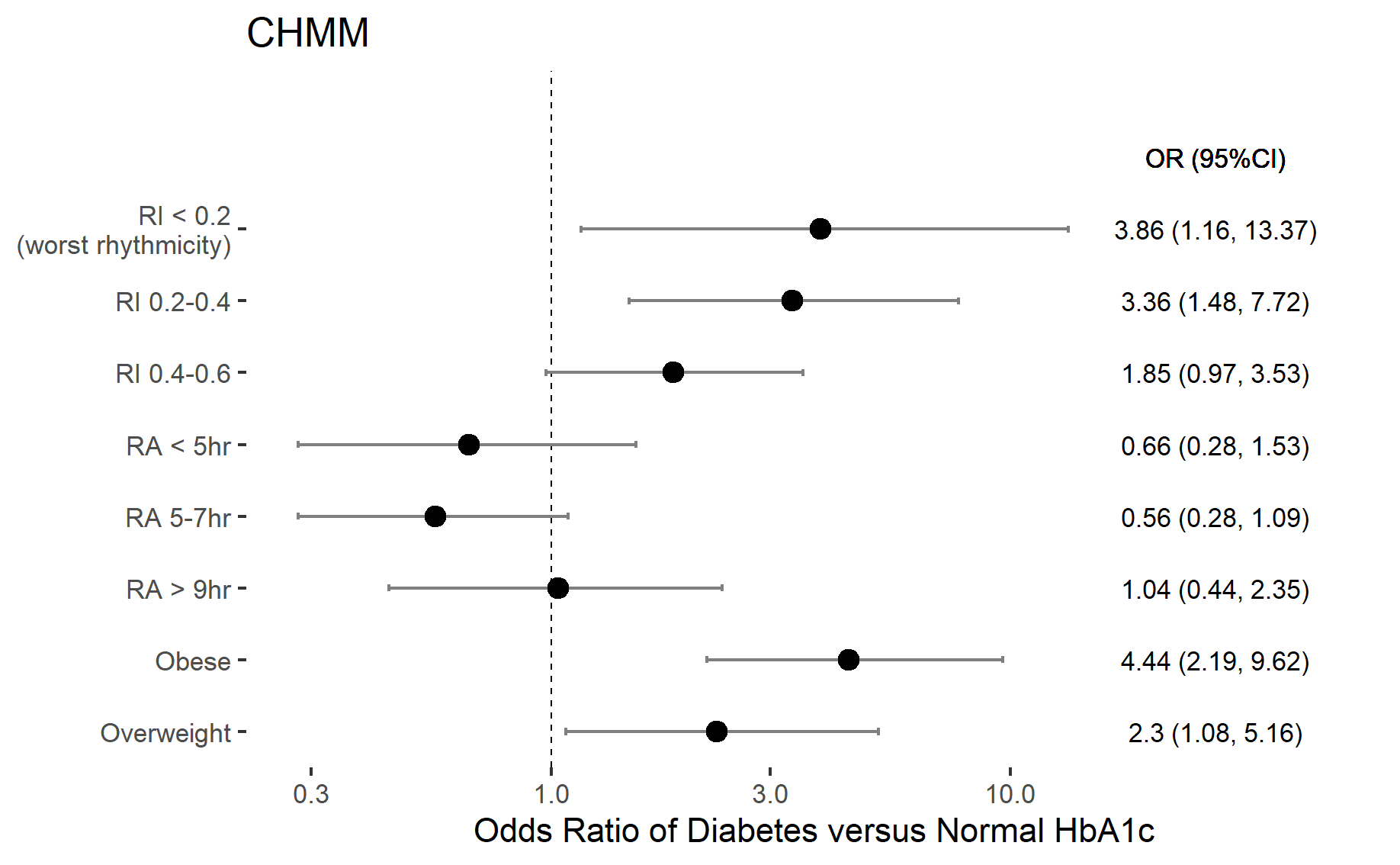} 
    \caption{Association between diabetic status and rest-activity rhythmicity using CHMM estimates. Results of odds ratios are presented, estimated from the logistic regression model, adjusting for age, gender, weight status characterized by BMI. Increased odds of having diabetes versus normal HbA1c levels is monotonously associated with lower values of RI, i.e., worse overall rhythmicity.}
    \label{fig:odds_ratios_nhanes_chmm}

\end{figure}

\end{document}